\let\texyear\year
\documentclass{ieeeaccess}

\usepackage{cite}
\usepackage{amsmath,amssymb,amsfonts}
\usepackage{algorithmic}
\usepackage{graphicx}
\usepackage{textcomp}

\let\ieeeaccessyear\year
\let\year\texyear

\usepackage{subcaption}
\usepackage{adjustbox}
\usepackage[font=footnotesize]{caption}
\usepackage{microtype}
\usepackage{setspace}			
\usepackage{multicol}			
\usepackage{multirow,bigdelim}
\usepackage{booktabs}
\usepackage[table]{xcolor}    
\usepackage{ellipsis}			
\usepackage{enumerate}
\usepackage{textcomp}
\usepackage{mathptmx}
\usepackage{mathrsfs}
\usepackage{tikz}
\usepackage{pgfplots}
\tikzset{every picture/.style={font issue=\footnotesize}, font issue/.style={execute at begin picture={#1\selectfont}}}
\usepackage{placeins}

\let\year\ieeeaccessyear
\definecolor{accessblue}{RGB}{0,105,154}

\def\BibTeX{{\rm B\kern-.05em{\sc i\kern-.025em b}\kern-.08em
T\kern-.1667em\lower.7ex\hbox{E}\kern-.125emX}}
\begin{document}
\history{Date of publication xxxx 00, 0000, date of current version xxxx 00, 0000.}
\doi{10.1109/ACCESS.2017.DOI}

\title{Determining Distributions of Security Means for WSNs based on the Model of a Neighbourhood Watch}
\author{\uppercase{Benjamin F\"orster}\authorrefmark{1}, \uppercase{Peter Langend\"orfer\authorrefmark{2}, and Thomas Hinze}.\authorrefmark{3}}
\address[1]{Wireless Systems, Innovations for High Performance Microelectronics, Im Technologiepark 25, Frankfurt (Oder), Germany (e-mail: bfoerster@ihp-microelectronics.com)}
\address[2]{Wireless Systems, Innovations for High Performance Microelectronics, Im Technologiepark 25, Frankfurt (Oder), Germany (e-mail: langendoerfer@ihp-microelectronics.com)}
\address[3]{Faculty of Biological Sciences, Friedrich Schiller University Jena, Ernst-Abbe-Platz 2, Jena, Germany (e-mail: thomas.hinze@uni-jena.de)}

\markboth
{Förster \headeretal: Determining Distributions of Security Means for WSNs based on the Model of a Neighbourhood Watch}
{Förster \headeretal: Determining Distributions of Security Means for WSNs based on the Model of a Neighbourhood Watch}

\corresp{Corresponding author: Benjamin Förster (e-mail: bfoerster@ihp-microelectronics.com).}

\begin{abstract}
	Neighbourhood watch is a concept allowing a community to distribute a complex security task in between all members.
	Members carry out security tasks in a distributed and cooperative manner ensuring their mutual security and reducing the individual workload while increasing the overall security of the community.
	Wireless sensor networks (WSNs) are composed of resource-constraint independent battery driven computers as nodes communicating wirelessly.
	Security in WSNs is essential to prevent attackers from eavesdropping, tampering monitoring results or denying critical nodes from providing their services and potentially cutting off larger network parts.
	The resource-constraint nature of sensor nodes prevents them from running full-fledged security protocols.
	Instead, it is necessary to assess the most significant security threats and implement specialised security solutions.
	A neighbourhood watch inspired distributed security scheme for WSNs has been introduced by Langendörfer aiming to increase the variety of attacks a WSN can fend off.
	The framework intends to statically distribute requirement-based selections of online security means intended to cooperate in close proximity on large-scale static homogeneous WSNs.
	A framework of such complexity has to be designed in multiple steps.
	We determine suitable distributions of security means based on graph partitioning concepts.
	The partitioning algorithms we provide are NP-hard.
	To evaluate their computability, we implement them as $0-1$ linear programs (LPs) and test them on WSN models generated with our novel $\lambda$-precision unit disk graph (UDG) generator.
\end{abstract}


\titlepgskip=-15pt

\maketitle

\section{Introduction}\label{sec:intro}
WSNs are networks consisting of independent low power computing units called sensor nodes running on battery, communicating wirelessly and carrying out monitoring or controlling tasks. 
Information gathered by sensor nodes is transmitted to base stations (BSs).
In large-scale static homogeneous WSNs considered in this work, the communication takes place hop-by-hop.
Additionally, only a small subset of nodes is connected to a BS.
The term {\itshape static} means nodes in the network are immobile and placed at a fixed position.
{\itshape Homogeneous} implies that all nodes in the network have the same hardware capabilities.
In large-scale WSNs, information to and from nodes is transmitted via intermediate nodes (hop-by-hop). 
Especially, when applied to critical infrastructures, WSNs need to ensure certain security attributes regarding transmitted data.
In general, WSNs are vulnerable to a multitude of attacks.
Therefore, security risks have to be well assessed and covered in the design of the network.
The limited computational power and energy supply of nodes constrain the types, complexity and scoop of security means suitable to WSNs.
Hence, we have to compromise between security and longevity of a WSN.
Such a compromise requires the necessity to identify the most likely and costly threats to a WSN and select security means accordingly.
A number of concepts to identify and priorities security means based on numerous properties have been proposed \cite{ki2017requirements,jiang2012attack}.
The general risk assessment, independent of the techniques that come along with it, comprises the steps: {\itshape risk identification}, {\itshape analysis} and {\itshape evaluation} \cite{ki2017requirements}.
The {\itshape risk identification} assess a system to identify external threats and system vulnerabilities.
The {\itshape risk analysis} assess the likeliness of identified risks to be exploited and the resulting consequences.
Finally, the {\itshape risk evaluation} derives consequential actions based on the analysis results.
One well established concept to do so are {\itshape attack defence trees} \cite{garg2019tree,jiang2012attack,langend2019security}.
An attack tree based risk assessment approach for location-privacy in WSNs is presented in \cite{jiang2012attack}.
In \cite{garg2019tree} an attack defence tree based risk assessment model for unmanned aerial vehicles model is researched.
Attack-defence trees are graphical models illustrating potential attack scenarios and corresponding countermeasures for a system and evaluate how various attacks can be mitigated through countermeasures and how those are interrelated.
In \cite{langend2019security}, Langendörfer proposed an extended concept of attack defence trees considering the resource limitations of the underlying nodes called ``Attack Defence Resource Trees'' (ADRT).
It targets a selection of security means based on pre-defined security incentives tailored to the area of application of a WSN and the resulting most likely threat scenarios while taking into account the hardware constraints and longevity of devices (resources).
Even with an optimal selection of security means, the coverage of a larger scope of threat scenarios is limited.

Hence, \cite{langend2019security} further proposed the concept of a neighbourhood watch-inspired in-network security. 
It assumes that an optimal selection of security means properly distributed throughout wirelessly communicating resource-constraint embedded devices, suitable to collaborate increases the threat coverage while keeping the individual security tasks load per node manageable. 
Taking into account the cooperation of security means, ADRTs are further extended by a cooperative component to ``Cooperation-based Attack Defence Resource Trees'' in \cite{langend2019security}. 
Collaborative security schemes coordinate nodes for more advanced threat detection and mitigation.
When referring to either of those two terms, we encompass the broader set including both cooperative and collaborative security approaches.
Existing collaborative security frameworks for pervasive systems like WSNs are specialised to very specific system constraints, sizes, network topologies, protocols and threat scenarios.
While those frameworks \cite{bhushan2016concealed,valero2012di,saxena2010dsf} are capable to offer a high degree of security and an increased threat coverage, they are not applicable to a wide range of WSNs.
Those frameworks go beyond the scope of collaborative distributed intrusion detection systems (IDSs) \cite{sharma2019survey,li2021surveying,du2021optimal} by integrating concepts of intrusion detection, prevention and complex communication strategies.
It follows that each new WSN requires the development of a new framework, and this is preceded by corresponding research work.
To achieve timely, cost-efficient, adaptable and reusable cooperative security solutions for a wide array of WSNs a different approach is necessary.
Based on system properties and constraints, the area of application, security and lifetime requirements a selection of components has to be determined.
Determining a proper security configuration for given requirements and constraints can be set up as a design space exploration (DSE).
There has been a lot of research regarding DSE models for embedded systems including WSNs focussing i.a. on aspects like security, safety, longevity and network topology (with regards to the placement of nodes) \cite{peter2008network,cionca2009tool,peter2011tool,peter2012tool,cionca2012configuration,lange2014tool,eiben2015introduction,peter2015component,kirov2017archex,kirov2018optimized,nuzzo2019optimized,tsiskaridze2021automating}.
Those concepts attempt to determine optimal configurations of components and subcomponents taking into account their interplay of a system with regards to a set of requirements and constraints.
Research in the domain of WSN security is constrained in terms of exploring the intricate interrelationships within the design space, particularly when it involves diverse forms of interaction and collaboration arising from hierarchical or clustered network structures.

The design of WSNs encompasses a wide array of factors represented by variables.
For the practicality of DSE for the configuration of different aspects of WSNs, it is crucial to narrow down the variables to a computationally feasible subset.
Resulting design spaces are expected to encompass interdependent parameters.
Moreover, exploration models often contain non-linear non-convex constraints and objective functions.
Conflicting objectives such as longevity and highest possible security standards can be handled with a multi-objective optimisation resulting in a Pareto-efficient solution space.
In order to enhance DSE approaches by incorporating the interplay of security means and various operational scopes (such as clusters or hierarchical structures), the optimisation models often exhibit a multitude of variables and continue to exacerbate the difficulty that comes with non-convex and non-linear characteristics.
Therefore, only a limited number of aspects can be considered in the optimisation process to still achieve a satisfying solution.

In this publication, we propose static distribution concepts of fix numbers of security mean types intended to cooperate by partitioning the WSN accordingly.
The approach intends to distribute different security means in the WSN with the objective to ensure the availability within the range of each node.
It provides a generic solution for distributed/cooperative security configurations in large-scale static homogeneous WSNs. 
The availability of security means in proximity of each node is a prerequisite for the neighbourhood watch inspired security framework.
It ensures short communication paths between nodes contributing with different security means to common security requirements. 
The partition concepts we are going to introduce offer a high availability of different security mean types in proximity of nodes in the WSN.
Such a partitioning facilitates multiple associations per node, well-suited for in-network cooperation and despite a static placement of nodes and static distribution of security means a flexible load balancing.
Therefore, it provides a generic solution suitable for a cooperative approach of a neighbourhood watch inspired security concept achieving timely, reliable, and energy-efficient collaborative threat prevention, detection and handling. 
To create a DSE approach that allows requirement-based security configurations for WSNs (e.g. by adapting existing concepts \cite{kirov2017archex}), our partitioning concept builds a foundation that improves the computability by limiting the number of model variables. 
In order to highlight the limitations of a concept that statically distributes security means in a large-scale WSN, we make a number of assumptions.
A fundamental premise for the success of a framework employing static security mean distributions is the assumption that an attacker possesses no insider-level knowledge about the distribution. 
Further, we assume that a trusted communication between sensor nodes has been established and as already mentioned, that the WSN is static (immobile nodes).
There are three scenarios of security means distributions we are going to evaluate: a single security mean per node, a fixed number of security means per node, a load-based distribution of a variable number of security means per node.
For the latter one, it is necessary to pinpoint a common resource capacity per node for all nodes available for security tasks and a resource requirements for all considered security means.
To ascertain the distributions, we model the WSNs as undirected graphs wherein the nodes symbolise sensor nodes.
The edges of the graphs indicate the connectivity between sensor nodes within transmission range of the underlying WSN.
In order to distribute the security means, we determine $0-1$ LPs to compute suitable optimal graph partitions. 
Optimal with regards to our model and the defined objective function.
The $0-1$ LPs fall into the complexity class of non-deterministic polynomial time (NP) hard problems \cite{karp1972reducibility}.
Therefore, it is imperative to empirically assess whether an optimal solution falls within the feasible and efficiently computable limits of our input sizes for numerous WSNs with realistic topologies and node quantities.
Network sizes of WSNs with $20$ up to $300$ nodes have been evaluated.
The graphs representing the WSNs for the evaluation process have been generated as {\itshape random $\lambda$-precision unit disk graphs} (UDGs). 
A UDG is an undirected geometric graph in which each node has a fixed position in euclidean space and two nodes have a common edge if their distance is below a fixed threshold (transmission range) common for all nodes. 
A $\lambda$-precision graph is a geometric graph in which all pairs of nodes are at least $\lambda$ apart. 
To generate desired graph topologies, we provide a table of generator seeds for combinations of node numbers, desired average node degrees and covered generation plane space. 
Seeds are the input values for the generator that are likely to result in random graphs with desired properties.
The average node degree is the arithmetic mean of edges connected to each node for all nodes in a graph.
The generation plane is in our context a unit square in which the nodes of our random graphs are distributed.
The generator is written in Python and utilises the NetworkX library \cite{hagberg2008exploring} to some degree. 
It allows to create graphs with an even degree distribution and a low variance of the {\itshape local cluster coefficient} controllable via $\lambda$. 
The local cluster coefficient is a measure indicating the connectivity of the neighbourhood of a node.
The generator allows further manipulations of graph properties like enabling to enforce a desired average node degree and receiving connected bridge-free graphs.
The $0-1$ LPs have been evaluated using Python with Pyomo \cite{hart2011pyomo} and Gurobi \cite{gurobi} to model and solve the linear optimisation problems partitioning the graphs for an optimal distribution of security means. 

Cooperation of nodes increases their load.
In a cooperation a node has to providing different services and handle requests. 
Such communication overhead increase further if a larger number of nodes direct their requests to a single node.
An equal distribution of security means across the WSN balances the load and increases the availability of neighbouring nodes offering specific services.
The necessity for load balancing is caused by imbalances as result of topologically conditioned unequal distributions and routing changes due to various reasons.
Due to our partitioning scheme a certain likeliness of alternative nodes providing the same service or accomplishing the same task in proximity of a node is given.
The association of nodes and their cooperation partners will not be taken into account when determining the distribution.
Rather it will be dynamically determined by the nodes themselves.

In Section \ref{sec:background}, we acquaint the reader with mathematical terminology and definitions essential for comprehending the paper.
Section \ref{sec:rel_works} delves into the current state of distributed security solutions, dominating sets, domatic partitions, and graph generators specifically tailored for large-scale static homogeneous WSNs.
Following that, in Sections \ref{sec:distr_sec} and \ref{sec:soft_dom}, we illustrate the concept of graph partitioning within the context of the neighbourhood watch inspired security scheme introduced in \cite{langend2019security}.
Subsequently, in Section \ref{sec:udg_gen}, we introduce a $\lambda$-precision UDG generator designed for large-scale static homogeneous WSNs.
In Section \ref{sec:empirical_setup}, we familiarise the reader with the experimental setup used to evaluate the feasibility of the proposed graph partitioning concepts, which have been formulated as $0-1$ linear programs (LPs) and computed on the $\lambda$-precision UDGs generated by our novel graph generator.
Finally, we present and analyse the test results in Section \ref{sec:res_eval} and draw conclusions regarding various accomplishments of our paper in Section \ref{sec:conclusion}.

\section{Background}\label{sec:background}
We introduce mathematical terms and definitions related to graph theory and mathematical optimisation as well as terms necessary for the empirical evaluation.\\

{\bfseries Cardinality of Sets:} The cardinality of a set indicates the number of elements a set contains notated as follows $|\{\cdot\}|$.\\

{\bfseries Undirected Graph:} 
An undirected irreflexive graph $G=(V,E)$ is defined as a finite set of nodes $V$ and a set of edges:
\begin{equation}\label{eq:undirected}
	E\subseteq \{\{v,w\}|v,w\in V\wedge v\neq w\}
\end{equation}
Throughout this work, we exclusively utilise undirected and irreflexive graphs.\\

{\bfseries Subgraph:} A subgraph of an undirected graph $G=(V, E)$ is defined as $SG=(V', E')$ with $V'\subseteq V$ and $E'\subseteq E$ with $\forall\{v,w\}\in E':v,w\in V'$.\\

{\bfseries Connected Graph:} An undirected graph is connected when there are no two nodes in the graph without a path.\\

{\bfseries Connected Component:} In an undirected graph, a connected component is a connected subgraph that is not part of any larger connected subgraph.\\

{\bfseries Bridge:} In an undirected graph consisting of $c\in \mathbb{N}_{>0}$ connected components, a bridge is an edge, whose absence decomposes it into $c+1$ connected components.\\

{\bfseries Bridge Path:} In an undirected graph $G = (V, E)$, there is a bridge path between nodes $u$ and $v$ iff there is a unique cycle-free path $P$ exclusively composed by a sequence of {\itshape bridges} over a subset of nodes from $V\setminus\{u,v\}$ connecting $u$ with $v$ in which all contained nodes except $u$ and $v$ have a node degree of two and it does not exist any longer path $Q$ with the same properties containing $P$.\\

\sloppy{\bfseries Geometric Graph:} A geometric graph is an undirected graph in a $d$-dimensional metric space $[0, 1)^d$ and edges are added based on their pairwise distance $r_\text{tr}$ (transmission range) determined by a defined distance function. 
The distance $r_\text{tr}$ in a geometric graph is fix for all nodes and node pairs of the graph. 
Throughout this work, we always refer to this distance as $r_\text{tr}$.\\

{\bfseries Random Geometric Graph:} A random geometric graph (RGG) is a geometric graph in which nodes are placed randomly.\\

{\bfseries Unit Disk Graph:} A unit disk graph (UDG) is a geometric graph in a two-dimensional euclidean space with an euclidean distance metric applied to them.\\

{\bfseries $\lambda$-precision Graph:} A $\lambda$-precision graph is a geometric graph in which the minimal distance between each pair of nodes is at least $\lambda$.\\

{\bfseries Neighbourship Function:} We define the neighbourship of a node $v$ in an undirected graph $G=(V,E)$ with $v, w\in V$ as follows:
\begin{equation}
	N[v]:=\{w|\{v, w\}\in E\}\cup\{v\}
\end{equation}
$\;$

{\bfseries Node Degree:} A node degree of a node $v\in V$ of an undirected graph $G=(V,E)$ is the number of edges of the graph the node participates in:
\begin{equation}
	\deg[v]=\left|\{e|\forall e\in E:\,v\in e\}\right|
\end{equation}
$\;$

{\bfseries Average Node Degree:} The average node degree of an undirected graph $G=(V,E)$ is the arithmetic mean of the node degree of each node in the graph relative to the number of nodes as follows:
\begin{equation}
	\deg_\text{avg}[G]=\sum_{v\in V}\frac{\deg[v]}{|V|}
\end{equation}
$\;$

{\bfseries Local Cluster Coefficient:} The local cluster coefficient is a measure indicating how well the neighbourhood of a node is connected. 
Following \cite{watts1998collective}, the local clustering coefficient for undirected graphs is defined as:
\begin{equation}
	C[v]=\frac{2\cdot\left|\{e| e\in E \wedge e=\{w,u\} \wedge w,u\in N[v]\backslash\{v\}\}\right|}{|N[v]\backslash\{v\}|\cdot(|N[v]\backslash\{v\}|-1)} 
\end{equation}
$\;$

{\bfseries Variance of the Node Degree Distribution:} We define the variance of the node degree distribution for a RGG $G=(V,E)$ as follows:
\begin{equation}
	\text{Var}_{\deg}[G]=\sum_{v\in V}\frac{\left(\deg[v]-\deg_\text{avg}[G]\right)^2}{|V|}
\end{equation}
$\;$

{\bfseries Linear Program:} A LP or linear optimisation is a method which tries to optimise a mathematical model based on linear relationships with the following standard form:
\begin{equation}\label{eq:sf_lp}
	\begin{array}{llll}
	\max        & \mathbf{c}^T\cdot \mathbf{x}	&       		& \rdelim\}{1}{*}[objective function]\\[4pt]
\text{s.t.} & \mathbf{A}\cdot \mathbf{x}	&\leq \mathbf{b}& \rdelim\}{2}{*}[constraints] \\
			& \mathbf{x}					&\geq \mathbf{0}& 
\end{array}
\end{equation}
with the vectors $\mathbf{b}$ and $\mathbf{c}$ and with a matrix $\mathbf{A}$ that have to be known to the problem. The vector $\mathbf{x}$ contains the variables whose values have been optimised. Linear programs are called in this way because the objective function as well as the equality and inequality constraints are linear.

In a {\itshape 0-1 linear program}, the components of the vector of variables $\mathbf{x}$ is bound to $\{0,1\}$. For integer linear programming as well as 0-1 linear programming without objective function it is known that they belong to the class of NP complete problems \cite{karp1972reducibility}. With objective function, their complexity is not bound to an upper limit and the problems are therefore considered to be NP hard. However, experience has shown that 0-1 linear programs perform better than integer linear problems even when they rely on significantly more variables.\\

\begin{figure}[t]
	\centering
	\includegraphics[width=0.4\textwidth]{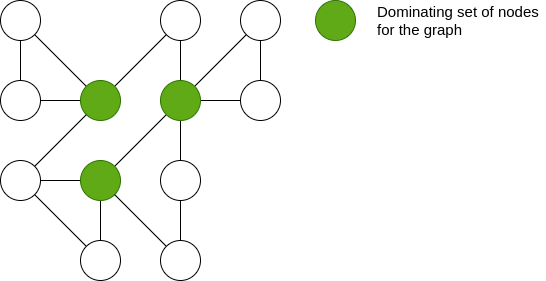}
	\caption[Dominating Set]{The set of green nodes is a dominating set in the given graph.}
	\label{fig:dominating_set}
\end{figure}

{\bfseries Dominating Set:} A dominating set $D$ is a set of nodes of an undirected graph $G=(V,E)$ for which holds: 
\begin{equation}
	D\subseteq V\,\,\textrm{whereas}\,\,\forall v\in V:\;D\cap N[v]\neq\emptyset 
\end{equation}
In Fig. \ref{fig:dominating_set}, an example for a dominating set of nodes for a graph is given. As the definition implies, every node in this graph is either part of the dominating set or adjacent to a node from the set.\\

\begin{figure}[t]
	\centering
	\includegraphics[width=0.4\textwidth]{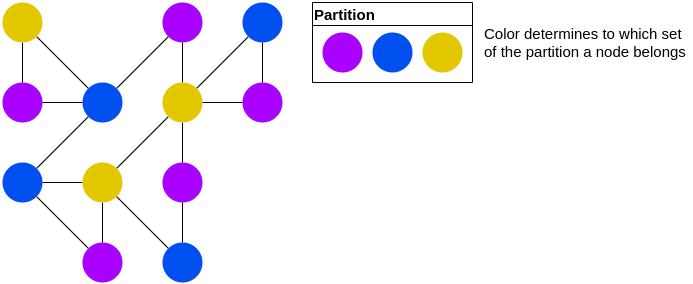}
	\caption[Domatic Partition]{The example shows a graph in which the nodes are mapped to a domatic partition consisting of three dominating sets.}
	\label{fig:domatic_partition}
\end{figure}
\begin{figure}[t]
	\centering
	\includegraphics[width=0.4\textwidth]{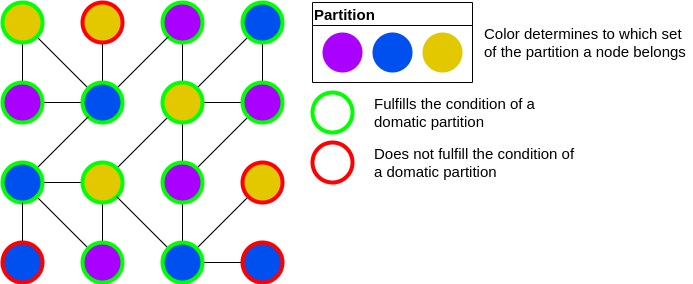}
	\caption[Counterexample Domatic Partition]{The partition of the graph is not a domatic partition because there exist nodes in at least one dominating set of the partition that has no neighbourship with at least one node of each of the other dominating sets of the partition.}
	\label{fig:domatic_partition_error}
\end{figure}

{\bfseries Domatic Partition:} A domatic partition $\mathbb{D}(G)$
is a decomposition of nodes $V$ of a graph $G=(V,E)$ into disjoint dominating sets with:
\begin{equation}\label{eq:domatic_cond}
	\bigcup_{D\in \mathbb{D}}D=V\;\wedge \bigcup_{\substack{D_1, D_2\in \mathbb{D} \\ D_1\neq D_2}}D_1\cap D_2=\emptyset
\end{equation}
A domatic partition can also be defined using the neighbourship term of graphs. 
Then, a set of dominating sets in $G$
\begin{equation}
	\mathbb{D}(G)=\{D|D\subseteq V,\;\forall v\in V:\;D\cap N[v]\neq\emptyset\}
\end{equation}
is a domatic partition iff Equation \eqref{eq:domatic_cond} holds. 
We define a $n$-domatic partition as a partition of $G$ into $n$ disjoint dominating sets. An example can be seen in Fig. \ref{fig:domatic_partition}. When referring to a node satisfying the properties of a domatic partition, the set consisting of the node itself and its adjacent nodes have to have a non-empty intersection with all sets of the domatic partition:
\begin{equation}
	v\in V:\forall D\in \mathbb{D} :N[v]\cap D\neq\emptyset
\end{equation}
In Fig. \ref{fig:domatic_partition_error}, we provide an example for a partition in which a number of nodes does not satisfy the definition of a domatic partition.
A domatic partition of a WSN ensures that each sensor node has at least one direct neighbour of each dominating set of the partition or is a member of the set. The size of the domatic partition is given by the number of different security means that have been applied to the network. All sensor nodes in the same dominating set of the partition implement the same security mean.
In case all nodes in the same dominating set implement the same security mean, we achieve a distribution of security means in which all nodes either implement a security mean or are directly adjacent to a node that does. Therefore, the set of sensor nodes and its neighbours have no empty intersection with any of the sets of the partition. Hence, all security means applied to the WSN are present in the neighbourship of each node.

\section{Related Works}\label{sec:rel_works}
In \cite{langend2019security} a neighbourhood watch inspired concept for a cooperative distributed static security framework has been introduced. 
The objective of distributed security solutions is to cover a wider range of threat scenarios in a large-scale static homogeneous WSN.
This section is divided into three parts. 
The first subsection explores research work towards distributed security solutions for WSNs. 
The second part evaluates existing research regarding dominating sets and domatic partitions. 
In the third subsection, we discuss graph generators as model for WSNs.

\subsection{Distributed Security Solutions for WSNs}
A number of publications propose cooperating security means for WSNs that provide mutual protection.
The paper \cite{saxena2010dsf} introduces a security framework concept for static heterogeneous WSNs. 
Each set of nodes is assigned to a cluster head (CH) (a more powerful sensor node). 
Nodes running IDSs notify their associated CH about identified threats or CH are informed by CHs in close proximity. 
If a threat is detected and communicated to a CH it will be propagated to other CHs in the WSN. 
Clusters that consider the threat imminent for their own cluster react by redistributing security means on associated nodes based on the threat scenario. 
Therefore, the CH holds a set of security means which can be implemented on or revoked from the sensor nodes. 
This allows a dynamic threat evaluation and flexible reactions. 
The proposed security framework for static heterogeneous WSNs \cite{saxena2010dsf} has been tested in a simulation including a network with $2000$ regular nodes and $10$ gateway nodes. 
The energy consumption was only evaluated for regular nodes, for CHs it was considered unlimited. 
To test the simulated sensor network, seven abstract attack patterns have been implemented and for each scenario $200$ sequential attacks have been executed. 
The authors of \cite{saxena2010dsf} evaluated the simulation based on two metrics, the success rate (number of nodes alive after an attack) and the energy consumption (average percentage of energy of all surviving nodes).
For comparison, WSNs implementing one security mean or multiple static security scheme frameworks have been used.
The results show that the proposed framework provides the highest success rate while also consuming the highest amount of energy in each simulation. 
The contribution \cite{valero2012di} presents a security framework that has been developed and implemented on a real WSN based on \cite{saxena2010dsf}. 
The test of the resulting security framework has been executed on a rather small WSN with only six nodes. One node acted as the CH which communicated directly to a base station. 
The authors assumed two kinds of attack scenarios. 
One in which only a single kind of attack is started on the WSN and one in which two kinds of attack are launched in succession. 
The results show that the WSN implementing the framework was able to recover from all tested attacks even when they have been executed successively. 
The energy consumption has not been considered. 
Both papers propose a security solution with distributed security means for heterogeneous WSNs with rather powerful CHs. 
The heterogeneity of the WSN is not utilised by the frameworks to which the proposed one is compared.
The according statement from \cite{valero2012di} has very limited meaningfulness due to their limitation in executed test scenarios, measured parameters and small network size.

Another cooperative security solution is proposed in \cite{bhushan2016concealed}.
The paper proposes a concept to efficiently combine in-network intrusion detection and concealed data aggregation.
To do so, it utilises clusters.
In each cluster a CH is elected.
A CH fulfils multiple roles: it collects the data from the nodes in its cluster, runs intrusion detection on the data, aggregates them and finally forwards them hop-by-hop to a BS.


IDSs are distinguishable by many criteria \cite{butun2013survey}.
Whether the intrusion detection is executed online (in-network), offline (on BSs, external System/Server) or hybrid states if certain tasks are performed on nodes or on a centralised base station affecting whether a timely reaction is possible.
The choice of whether to execute intrusion detection online (within the network), offline (on BSs or external systems/servers), or in a hybrid manner depends on whether certain tasks are performed on nodes or BSs, which in turn affects the feasibility of timely response.
Based on their detection strategy, IDSs can be classified as anomaly-based, signature-based or hybrid.
\cite{stetsko2010neighbor} introduces a distributed neighbour based IDS. 
Each node monitors a set of neighbouring nodes by storing their attribute vectors sending warnings to other nodes in case a malicious anomaly has been recognised. 
If a number of nodes communicate the same anomaly, the network acts accordingly.
There are distinctions based on the intrusion and intruder type and so on.
A comprehensive overview of the classification IDSs is provided in \cite{butun2013survey}.
The publication \cite{stetsko2010neighbor} is built upon \cite{liu2007insider} which describes a similar distributed approach to detect misbehaving sensor nodes in local areas by comparing their behaviour vector with vectors from other direct neighbours.
Another popular concept is LiDeA \cite{krontiris2008lidea}. 
Nodes that detect irregularities in the network notify other close-by nodes to establish a vote. 
Notified nodes decide about the handling of the irregularity as well as the suspicious node. 
Therefore, nodes provide a number of modules that can be activated on demand and based on received information by broadcasting neighbours. 
Whether a node is assumed to be an intruder is determined based on a majority vote.
In \cite{riecker2015lightweight} a lightweight, energy-efficient IDS using mobile agents is introduced. 
These agents are sent through the network as regular messages and are temporally installed on addressed nodes.
Therefore, IDSs are dynamically distributed and instead of running on nodes permanently.
While agents are run by nodes, they collect information about their energy consumption and initiate warnings to the network if noticeable deviations are recognised. 
The transmission and installation of changing IDSs on nodes themselves, especially when executed on large-scale static homogeneous WSNs, significantly impacts the energy consumption.
Hence, it is assumed inappropriate for our subject of research.
Additionally, a IDS is intended not to introduce weaknesses into a WSN. 
The distribution of IDSs requires by itself a increased level of trust in the communication and sensor nodes.
However, many collaborative and distributed IDSs provide a promising basis to design a cooperative security framework integrating further components to collaborate.

Many of the concepts of cooperative/collaborative IDSs and security frameworks for WSNs are tailored to distinct properties, targeted application areas and satisfy specific security requirements.
One attempt to create a more universally applicable security concept is realisable using DSE. 
To do so, we can fall back on many existing concepts \cite{peter2008network,cionca2009tool,peter2011tool,peter2012tool,cionca2012configuration,lange2014tool,eiben2015introduction,peter2015component,kirov2017archex,kirov2018optimized,nuzzo2019optimized,tsiskaridze2021automating} and adapt them accordingly.
Therefore, a design space contains a number of components that can be combined to create a security framework for specific security requirements, lifetime expectancies and hardware constraints. 
This further necessitates metrics to assess the contribution of components to given requirements.
The proposed "collaborative attack defence resource trees" in \cite{langend2019security} is a concept displaying countermeasures and the attacks they prevent as well as the resulting resource costs to the WSN.
Further, the concept considers the weighting of the frequency of appearances of security means in a WSN based on likeliness and severity of attacks to the system.
For these concepts it is crucial to achieve an optimal distribution ensuring proximity between different types of security means and the corresponding nodes.

\subsection{Dominating Sets and Domatic Partitions}
To determine suitable static distributions of a fix number of security mean types intended to cooperate, local proximity is a key factor.
The concept of dominating sets and domatic partitions (alternatively $f$all $k$-colouring \cite{laskar2009fall}\footnote{A graph colouring problem that determines whether a graph can be coloured with $n$ colours so that in each node's neighbourhood all colours are present.}) is well suited. 
In a dominating set, each node is either adjacent to a node of the set or included in it. 
If sets represent security mean types, such a partition ensures the local proximity in a network.
Hence, a security mean type is either available on a selected node or a neighbouring node.
A domatic partition of a graph is the partitioning of it into disjoint dominating sets. 
If for a graph and a given number of security means such a partition exists, local proximity is ensured.
\cite{garey1979computers} states that the domatic partition problem that asks whether the nodes of a graph can be partitioned into $k\in\mathbb{N}_{\geq 3}$ dominating subsets is NP complete.
Known applications of dominating sets exist in the field of wakeup scheduling for WSNs \cite{floreen2007distributed,islam2009maximizing,misra2009efficient,mumey2013extending,pemmaraju2006energy,yu2014domatic}. 
However, in wakeup scheduling applications, dominating sets do not need to be disjoint. 
The major concern in energy-saving wakeup scheduling schemes is that at least one node in a neighbourhood of each node has to be kept awake to ensure that it can wakeup surrounding nodes. 
On the contrary, our applications require disjoint partitions into dominating sets.
The term {\itshape fractional domatic partition} was introduced in \cite{rall1990fractional}. 
This algorithm however determines a number of non-disjoint dominating sets.
Conversely, we attempt to determine a fixed size partition of disjoint sets approaching dominating sets as far as possible. 
Therefore, we approach the definition based on desired criteria, as we will introduce in the following sections.

In \cite{floreen2007distributed}, an approximation algorithm which tries to maximise the number of fractional domatic partitions in a graph to efficiently sleep schedule nodes is shown.
Furthermore, there exist a multitude of publications towards the domatic number and domatic partition problem with regards to different approximations and solution for specialised graph types providing lower and upper bound assumptions of their computational complexity. 
In \cite{feige2002approximating} a polynomial approximation algorithm estimating the lower and upper bounds of the domatic number on general graphs is presented. 
Additionally, \cite{feige2002approximating} determines a greedy approximation algorithm for domatic partitions of graphs. 
The algorithm computes as many small disjoint dominating sets as possible to receive a partition of fixed size. 
Other attempts achieving more precise bounds for the domatic number and domatic partition problem have been executed on  general graphs \cite{czygrinow2017improved,liang2012algorithmic} as well as special types of graphs, e.g. interval graphs \cite{rao1989linear} or RGGs \cite{mahjoub2010approximating}. 
\cite{pandit2009approximation} determines an approximation algorithm for domatic partitions on UDGs.
The survey \cite{yu2013connected} discusses and summarises a large number of research results and solutions towards different dominating set problems and compares the performances and properties of different algorithms proposed.

We intend to calculate our static distributions of security means (partitioning schemes) analytically.
The security means are distributed on nodes and not exchanged during runtime.
Therefore, an optimal distribution is a key factor for the overall performance of the security framework.
Furthermore, we have different requirements towards the partitioning compared to the distributions examined in sleep scheduling applications.

\subsection{Generators for Graph Models of WSNs}
A lot of research is done regarding the generation of graphs as model for different types of networks. 
One of the first models for generating random graphs as network model is the Erdős-Rényi model \cite{erdos1960evolution} expressed by $G(n,p)$. 
It is a popular way to construct Erdős-Rényi graphs.
In this model, $n$ labelled nodes are connected randomly. 
For all pairs of nodes, an edge is included with the probability $p$. 
Other popular models for random graph generators are the Barabási-Albert model \cite{barabasi1999emergence} and the Watts-Strogatz model \cite{watts1998collective}. 
The Barabási-Albert model aims to create scale-free graphs as network models. 
Therefore, the degree distribution in the resulting graphs follows a power law. 
The Watts-Strogats model generates graphs with small-world properties which are characterised by a high clustering coefficient and a low average shortest path length between nodes. 
\cite{kenniche2010random} reasons why RGGs are well suited as graph topology model for WSNs. 
In \cite{gilbert1961random} the author first mentions similar graph models called ``Random Plane Networks'' as representation of wireless networks. 
The resulting graphs are closely related to UDGs. 
Those type of graphs are most often the model of choice to represent WSNs. 
In \cite{jorgic2004localized} a model to generate WSNs that have a high probability to be connected as model for WSNs and ad hoc networks is introduced. 
To achieve the property {\itshape connected} with a high probability, the authors rely on a scheme that they call the proximity algorithm (PA). 
The PA places nodes iteratively on a finite plane. 
The first node is placed randomly within the generation plane. 
The following nodes are placed within radius $r$ of the previously placed nodes. 
Even so, $r$ is usually chosen larger than the distance in which two nodes are connected in a UDG, the likeliness of receiving a connected UDG using the PA increases significantly. 
A major downside of this approach is the likeliness for nodes in the graph to be highly clustered together.
One of the most popular concepts for the generation of random graphs as model for wireless ad hoc, actuator and wireless sensor networks has been published in \cite{onat2008generating}. 
The publication introduces two types of algorithms to generate random UDGs. 
Centre node based algorithms are one type and acceptance/rejection based algorithms are the other. 
With centre node based algorithms, a node out of the previously placed nodes is chosen (centre) and the new node is placed in reach of the chosen centre. 
The paper presents four different algorithms. 
Each of them introduces different centre choosing strategies. 
The second type, acceptance/rejection based algorithms, works by iteratively choosing random node locations. 
The selected location is accepted or rejected based on given constraints. 
The authors propose three different algorithms to apply the acceptance/rejection based concept. The resulting graphs are called {\itshape constrained connected random UDGs} (C-CRUG). 
The term {\itshape constrained} reflects the circumstance that the placement is not completely random but constrained by the node positions of previously placed nodes. 
Moreover, the term {\itshape connected} means that the final result will only be accepted if the graph is connected.

In \cite{onat2008generating} the authors relied on three different constraints.
The proximity constraint which is closely related to the PA by \cite{jorgic2004localized}. 
It ensures that each node is placed close to previously placed nodes increasing the likeliness for the resulting graph to be connected. 
Each node successive to the first node has to be placed within an approximated radius of previously placed nodes. 
The radius is estimated based on further desired graph properties. 
As with the PA, the radius constraining the node placement increases the likeliness of receiving islands of strongly clustered nodes.
The actual radius used to decide whether two nodes in the graph are connected is determined as the $\frac{N\cdot d_{\text{avg}}}{2}$th shortest edge with $N$ the number of nodes in the graph and $d_{\text{avg}}$ the average node degree. 
Therefore, resulting graphs are not guaranteed to be connected.
The second constraint used in \cite{onat2008generating} is the maximum degree constraint. 
It accepts the placement of a new node only if it does not increase the degree of the already placed nodes above a given maximum value. 
The third and last constraint was named the coverage constraint. 
With the coverage constraint, a new node location is only accepted if it extends the area that will be covered by the nodes of the graph sufficiently.
Regarding the proximity constraint \cite{onat2008generating}, a minimal distance in between nodes equal to the $\lambda$ in $\lambda$-precision graphs is considered. 
But the paper merely employs the distance to avoid that two nodes will be placed on the same coordinate instead of utilising $\lambda$ for a better spatial node distribution. 
Hence, proposed centre node based algorithms from \cite{onat2008generating} are:\\

{\bfseries Minimum Degree Proximity Algorithm (MIN-DPA):} 
It distributes nodes more uniformly while still maintaining connectivity. 
The first node is placed completely at random. 
Succeeding nodes are placed in the range of previously placed nodes with the lowest degree. 
In case there are multiple equally suitable contenders, all nodes get assigned a weighting scheme based on further criteria.\\ 

{\bfseries Clustered Minimum Degree Proximity Algorithm (C-MIN-DPA):} 
Instead of distributing homogeneous nodes, this algorithm starts to distribute access points (APs). 
They are assumed to be connected first.
The nodes will then be placed in close proximity to the APs, so they are connected to them.\\

{\bfseries Weighted Proximity Algorithm (WPA):} 
This algorithm is similar to MIN-DPA but it considers all previously placed nodes as centres instead of just the ones with the lowest degree. 
To randomly select nodes, all nodes associated with a weight relative to their node degree. 
Therefore, nodes with a higher degree receive a smaller weight.\\ 

{\bfseries Eligible Proximity Algorithm (EPA):} 
The nodes and their transmission ranges that serve as possible candidates for the location of the next node are selected by a given upper bound of the node degree. 
If the estimated node degree is larger than a given upper bound the placement of nodes is done according to WPA.\\

Proposed acceptance/rejection based algorithms are:\\

{\bfseries Maximum Degree Proximity Algorithm (MAX-DPA):} 
The algorithm sets a maximum degree constraint per node. 
A random node position is generated uniformly. 
If the node satisfies the proximity constraint as well as the maximum degree constraint the new position is accepted.\\

{\bfseries Coverage Algorithm 1 (CA1):}
The first node is placed completely at random. 
Subsequent nodes, choose a random coordinate. 
Their position is validated by a coverage constraint checking if the selected region is already sufficiently covered by previously placed nodes.\\

{\bfseries Coverage Algorithm 2 (CA2):} 
CA2 works similar to CA1 but with a stricter coverage constraint.
The covered portion of the sensing area for a new node location is explicitly computed with regard of the previously placed nodes. 
If the portion of the sensing area gained by the new node location is below a given threshold, the node location is rejected.\\

Our graph generator follows a different approach. 
We distribute nodes uniformly at random only constrained by a generation plane and a minimal distance in between nodes called $\lambda$-precision.
Instead of using $\lambda$ to prevent nodes from occupying the same spot as in \cite{onat2008generating}, we apply it to improve their spatial distribution and control a number of graph properties. 
When distributing a number of sensor nodes with fixed sensing range given by radius $r_\text{sensing}$, it is often of interest to maximise the monitored area. 

Therefore, $\lambda$ should usually be set between the radius $r_\text{sensing}$, a single sensor nodes sensing range and its transmission range $r_\text{tr}$ in which a sensor node is able to communicate. 
Choosing $\lambda$ larger than the transmission range prevents nodes from communicating. 
The rings resulting from the two radii $\lambda$ and $r_\text{tr}$ limit the maximum node degree of each node.
The choice of $\lambda$ and $r_\text{tr}$ relative to each other and relative to the generation plane determines the probability that a randomly generated graph is connected.

In general, the proposed generator is also suitable to be further developed into a topology generator allowing distributions of sensor nodes in target environments.
To extend our concept, we have to take into account the topological shape of landscapes as well as the varying transmission ranges based on different environmental conditions including various obstacles.
Even so, we distribute nodes in a unit square, the generation plane can have any shape.
To accommodate diverse landscapes and their environmental conditions, we can establish a connection between the $\lambda$-precision and the topological characteristics of specific areas, thereby enabling a higher concentration of sensor nodes in those regions.
Additionally, we demonstrate that even with the flexibility in node placement, it is possible to precisely adjust the local cluster coefficient and average node degree to meet specific requirements.

\section{Distribution of Security Means}\label{sec:distr_sec}
The neighbourhood watch inspired security concept \cite{langend2019security} intends to distribute different security means in a WSN enabling an increased threat coverage while keeping the energy consumption at bay.
In order to attain such a distribution, sensor nodes must establish a mutually beneficial cooperation among the applied security means.
Moreover, it is evident that nodes cannot continuously operate their security mean for neighbouring nodes as a service while ensuring their own longevity.
Instead, they can execute security means in specialised periodic patterns to detect malicious activity and subsequently process security violations.
The detection is solvable with a cooperative multilayer IDS approach, while the intrusion prevention pre-emptive and reactive requires further tools.
Suitable candidates are lightweight trust-and-reputation systems \cite{verma2017towards}, node isolation schemes \cite{ahmed2015survey}, resilient recovery techniques for compromised nodes \cite{strasser2006autonomous, hung2009attack}, lightweight encryption schemes \cite{hayouni2014survey}.
For the realisation of the security framework, following three assumptions have to be met:
\begin{itemize}
	\item[$\bullet$] trusted communication between sensor nodes has been established
	\item[$\bullet$] WSN is static (nodes are immobile)
	\item[$\bullet$] attacker has no knowledge regarding the distribution of security means
\end{itemize}
We consider static distributions of security means. 
Meaning, sensor node carry pre-installed security means and are incapable to exchange or rotate their security mean.
In the considered WSNs, we intend to distribute $n$ different types of security means.
Hence, we contemplate it mandatory to ensure the availability of each type of security mean in the neighbourhood of each node if possible. 
Therefore, nodes have access to all security mean types applied to the network. 
A distribution of this kind is achievable in case each security mean type is either implemented on the observed node or on one of its neighbours. 
Therefore, we aim to ensure that the set of all nodes implementing the same security mean type in union with the set of all neighbours of those nodes results in a set containing all nodes of the network.
Such a set is called a dominating set in graph theory.
Considering the set of nodes implementing the same security mean as a set for all security means, we get $n$ disjoint sets of nodes. 
Those sets are called dominating sets in graph theory. 
A partition of $n$ disjoint dominating sets of nodes of a graph is called a domatic partition. 
The number of applied security mean types distributed in a network implies the number of necessary dominating sets. 
The maximum number of disjoint dominating sets per graph is called domatic number $n$. 
Choosing $n$ larger than the domatic number of a graph, makes a partitioning in to $n$ disjoint dominating sets impossible.
Therefore, we introduce the term $n$-soft domatic partition.
An $n$-soft domatic partition attempts to compute a best possible fit as compromise with regards to the model parameters.
Another attempt to achieve an improved distribution of security means is the assumption to soften the neighbourhood term.
So far, we are considering direct neighbourhoods (one-hop).
Assuming multi-hop neighbourhoods, we are more likely to find an optimal partitioning as we later elaborate.
We also discuss fix and workload-based distributions of multiple security means per node.
Those approaches have currently limited practical applicability but can become relevant in the future.
The partition scheme, introduced in this work, we name {\itshape maximal/optimal $n$-soft domatic partition}. 
Primarily, we choose to focus on distributing one security mean per node are the resource limitations and longevity of nodes.
For this reason, we also focus on the one-hop neighbourhood in our analysis.
A one-hop neighbourhood significantly limits the number of nodes depending on a security mean type and therefore inflicting an increased load to it.
Viable alternative strategies are to consider multi-hop neighbourhoods allowing a more flexible rebalancing of node affiliations.

\section{Optimal and Maximal $n$ - Soft Domatic Partitions}\label{sec:soft_dom}
An $n$-soft domatic partition describes the partitioning of a graph into $n$ disjoint sets. 
While a domatic partition of size $n$ is restricted to graphs with a domatic number greater-equal to $n$, an $n$-soft domatic partition is computable for graphs with a domatic number lower than $n$.
We define two types of $n$-soft domatic partitions. 
Both types use different error terms to define either an {\itshape optimal} or a {\itshape maximal} $n$-soft domatic partition by minimising its respective error.
An $n$-soft domatic partition of size $n$ with nodes $V$ of a graph $G=(V,E)$ into disjoint sets of nodes $D_1,\ldots, D_n$ is defined as:
\begin{equation}
	\begin{split}
		& \mathbb{D}(G)=\\
		& \{D_i\subseteq V\;|\;i=1,\ldots,n\wedge\bigcup_{D\in\mathbb{D}}D=V\;\wedge\bigcup_{\substack{D_1, D_2\in\mathbb{D}\\D_1\neq D_2}}D_1\cap D_2=\emptyset\}
	\end{split}
\end{equation}
The definition of an $n$-soft domatic partition coincides with the definition of a regular partition of size $n$. 
After introduction of the terms optimal and maximal as additional conditions to the $n$-soft domatic partition, we define more specialised mathematical terms.\\

{\bfseries Optimal $n$-Soft Domatic Partition:} 
An $n$-soft domatic partition is called {\itshape optimal} iff {\itshape missing coverages} $e_\text{miss\_cov}$ from Equation \eqref{eq:miss_cov} is minimal.
In consequence, the optimal $n$-soft domatic partition minimises the sum of missing coverages over all nodes.\\

{\bfseries Maximal $n$-Soft Domatic Partition:} An $n$-soft domatic partition is {\itshape maximal} iff the number of {\itshape incompletely covered nodes} $e_\text{inc\_nodes}$ defined in Equation \eqref{eq:inc_nodes} is minimal.
Therefore, it is irrelevant whether $N[v]$ of a node $v\in V$ of graph $G=(V,E)$ has one or multiple non-empty intersections with any set $D\in\mathbb{D}$.\\

We use the newly introduced terms to determine a distribution of $n$ security mean types on sensor nodes of a WSN with a domatic number smaller than $n$. 
A maximal $n$-soft domatic partition ensures that the maximum number of nodes and its neighbourhood contains the full set of $n$ security means.
The optimal $n$-soft domatic partition guarantees that the number of missing coverages in a WSN is minimal. 
Hence, ensuring the sum of the absence of the number of security mean types in the inclusive neighbourhood of all nodes is minimal.\\

{\bfseries Error Terms in Soft Domatic Partitions:} The definition of optimal and maximal $n$-soft domatic partitions is based on two error terms. 
Those will be evaluated in our empirical analysis. 
The {\itshape missing coverages} are defined as the sum of the $n$ security mean types minus the security mean types present in the neighbourhood of a node $N[v]$ in a graph $G=(V,E)$ over all nodes $v\in V$: 
\begin{equation}\label{eq:miss_cov}
	\begin{split}
		& e_{\text{miss\_cov}} =\\ 
		&\quad\sum_{v\in V} \left(n - \left|\{D|\forall u\in N[v]:\exists D\in \mathbb{D}:\,D\cap u\neq\emptyset\}\right|\right)
	\end{split}
\end{equation}
with the set of nodes utilising the same security mean type creating a partition $D$ in the set of partitions $\mathbb{D}$.

In Fig. \ref{fig:domatic_partition_error}, we depict as example a graph with nodes of three colours {\itshape magenta}, {\itshape blue}, {\itshape yellow}. 
Each of those colours represents a set of nodes $D$ within a partition $\mathbb{D}$ of the given graph. 
All four nodes marked with a {\itshape red ring} contribute to the number of missing coverages. 
A node is {\itshape fully covered} if its inclusive neighbourhood contains nodes of all colours.
In Fig. \ref{fig:domatic_partition_error}, the number of missing coverages $e_\text{miss\_cov}$ is $6$.
There are four incompletely covered nodes surrounded by a red ring. 
The blue node at the lower left corner of the graph lacks the coverage of a yellow and a magenta security mean in its neighbourhood. 
So, its contribution to the coverage error is 2. 
The same holds for the blue node at the lower right corner of the graph. 
Here, two security means (yellow and magenta) are missing. 
The yellow node directly above has no access to the magenta security mean. 
Its coverage error is $1$. 
Finally, the yellow node marked with a red ring in the top line of the graph misses the magenta security mean. 
Resulting in a coverage error of $1$. 
In total, $e_\text{miss\_cov}$ results in $2+2+1+1=6$.

The second error term is named {\itshape incompletely covered nodes}. 
Counting the number of nodes $v\in V$ of $G=(V,E)$ for which the number of distinct security means in $N[v]$ is smaller than $n$:
\begin{align}
	& e_\text{inc\_nodes} = \nonumber\\
	& \quad\sum_{v\in V}f\left(n - \left|\{D|\forall u\in N[v]:\exists D\in \mathbb{D}:\,D\cap u\neq\emptyset\}\right|\right)\label{eq:inc_nodes}\\
	& \quad\text{with}\; f(x) = \begin{cases}
		0, & x < 1\\ 
		1, & x \geq 1
	\end{cases}\label{eq:f}
\end{align}
Let us again illustrate an example by the graph in Fig. \ref{fig:domatic_partition_error}. 
The four nodes marked with a red ring are incompletely covered. 
Hence, they are missing one or several distinctly coloured nodes in their inclusive neighbourhood.
In order to be completely covered by security means, a node needs to have access to all three colours (blue, magenta, yellow) within its direct neighbourhood. 
The error term $e_\text{inc\_nodes}$ identifies these nodes and sums up their occurrences. 
So, we obtain as result $e_\text{inc\_nodes}=4$.

In the worst case, for a graph $G=(V,E)$ with $V$ the set of nodes and $E$ the set of edges is at most
\begin{equation}\label{eq:max_inc_nodes}
	\max_{e_{\text{miss\_cov}}}(G) = \left|V\right| 
\end{equation}
incompletely covered nodes and 
\begin{equation}\label{eq:max_error}
	\max_{e_{\text{inc\_nodes}}}(G) = (n-1)\cdot\left|V\right|
\end{equation}
errors for a partition of size $n$, since each node has to be in at least one of the sets of the partition. 
An example for a worst case is the instance in which all nodes of a graph host the same security mean while the total number of required security means is higher $(n > 1)$. 

\subsection{Domatic Partition LP}
To compute the domatic partition of size $n$ of a given graph $G=(V,E)$, we conceptualise a $0-1$ LP without objective function.
The LP returns either a feasible solution or terminate with the response that no feasible solution exists.
In case a feasible solution exists, the assignments of the binary variables provide a feasible graph partitioning.
Hence, the $0-1$ LP determines a domatic partition of size $n$.

To construct a $0-1$ LP, we need to define a number of variables and construct a set of constraints representing the properties of a domatic partition. 
We define the variables $x^v_i\in\{0,1\}$ of the underlying $0-1$ LP. 
The upper index provides the identifier for the corresponding node $v\in V$ and the lower index links to the partition $i=1,\ldots,n$.
For each node $v\in V$, there are exactly $n$ variables, one for each partition. 
A value $1$ of a variable $x^v_i$ associates the node $v$ with the set $i$ of the partition. 
Otherwise, the value $0$ indicates that node $v$ is absent from the partition.

The first set of constraints we introduce ensures that each node has to be included in exactly one dominating set of the domatic partition:
\begin{equation}\label{eq:first_constraint}
	\forall v\in V:\sum_{i=1}^n x_i^v=1
\end{equation}

Moreover, we formalise that each node is either part of a dominating set or adjacent to one:
\begin{equation}\label{eq:dominating_set_constraint}
	\forall v\in V,\;\forall i\in\{1,\ldots,n\}: \sum_{w\in N[v]} x_i^w \geq 1
\end{equation}
Hence, for all dominating sets of a domatic partition the intersection with the set of adjacent neighbours $N[v]$ including the observed node $v$ is not empty.

The final $0-1$ LP without objective function reads as follows:
\begin{equation}\label{eq:sat}
	\begin{array}{lll}
		\forall v\in V:                              & \sum_{i=1}^n     & x_i^v = 1\\
		\forall v\in V,\;\forall i\in\{1,\ldots,n\}: & \sum_{w\in N[v]} & x_i^w \geq 1\\
		\forall v\in V,\;\forall i\in\{1,\ldots,n\}: &                  & x_i^v\in\{0,1\}
	\end{array}
\end{equation}
It determines whether a graph can be partitioned into an $n$-domatic partition. 
As a result it provides a domatic partition of the graph as solution.
Hence, the LP solves a satisfiability problem stating whether a given graph can be partitioned into $n$ disjoint dominating sets. 

We can extend the LP as proposed in the previous section by allowing each node to implement $k\in \mathbb{N}_{>0}$ different security means. 
To do so, it is only necessary to change the constraint from the Equation \eqref{eq:first_constraint} to:
\begin{equation}\label{eq:fract_dom}
	\forall v\in V:\sum_{i=1}^n x_i^v = k
\end{equation}
In the context of WSNs, the resulting partitioning yields a distribution of $n$ security mean types with $k$ security mean types implemented per node $v$ and all $v\in V:\;|N[v]|=n$ if one exists.

Furthermore, we can apply a variable number of security means per node based on an estimation of their respective resource costs. 
To do so, we apply fixed costs $m_i\in(0,1]$ to each security mean $i=1,\ldots,n$, a portion of the total available resources per node which w.l.o.g. is set to $1$. 
As long as the available resources on a node are not exhausted, additional security means can be applied.
The constraint from Equation \eqref{eq:first_constraint} is modified as follows:
\begin{equation}\label{eq:mixed_dom}
	\forall v\in V:\sum_{i=1}^n m_i\cdot x_i^v = 1
\end{equation}

\subsection{Optimal/Maximal $n$ - Soft Domatic Partition LPs}
Based on the LPs for the satisfiability conditions of domatic partitions from the preceding section, we introduce LPs for optimal and maximal $n$-soft domatic partitions.
At first, it is necessary to drop the constraints from Equation \eqref{eq:dominating_set_constraint}. 
The constraints ensure that each set of the partition is a dominating set. 
For maximal and optimal $n$-soft domatic partitions of graphs with $n$ greater than their domatic number, no partitioning into $n$ disjoint dominating sets exists. 
Instead, we introduce an objective function minimising either the number of missing coverages (Equation \eqref{eq:miss_cov}) or the number of incompletely covered nodes (Equation \eqref{eq:inc_nodes}) for optimal and maximal $n$-soft domatic partitions.

We start with Equation \eqref{eq:miss_cov} to minimise the missing coverages. 
Therefore, we transform the counting of missing coverages into a more applicable form for construction of partitions.
The objective function uses the previously defined function $f$ in Equation \eqref{eq:f}. 
All identifiers and variables such as $x_i^v$ and $n$ previously introduced in the Equations \eqref{eq:first_constraint} to \eqref{eq:sat} of the preceding subsection keep their semantics.
The LP to determine an optimal $n$-soft domatic partition then reads as follows:
\begin{equation}\label{eq:first_lp}
	\begin{array}{lll}
		\max         									& \sum_{v\in V}\sum_{i=1}^n f\left(\sum_{w\in N[v]} x_i^w\right) 	&\\[4pt]
		\text{s.t. } 									& \forall v\in V: 													&	\sum_{i=1}^n x_i^v=1\\
														& \forall v\in V,\;\forall i\in\{1,\ldots,n\}:	& x_i^v\in\{0,1\}
	\end{array}
\end{equation}
At first, we look at $\sum_{w\in N[v]}x_i^w$. 
The sum iterates over all $w\in N[v]$. 
It checks for each node $v$ associated with set $i$ whether a node of $N[v]$ is included in the set $i$ of the partition. 
The result is passed on to the function $f$ from Equation \eqref{eq:f}. 
The function indicates whether {\itshape at least one member of $N[v]$ is linked to set $i$} or {\itshape no member of $N[v]$ is included in set $i$} with the values $1$ and $0$ respectively.
Hence, the appearance of more than one node in $N[v]$ included in the set $i$ of the partition does not influence the optimisation result. 
The outer sums $\sum_{v\in V}\sum_{i=1}^nf\left(\sum_{w\in N[v]}x_i^w\right)$ ensure that the value is determined for all combinations of nodes $v\in V$ and sets $i$ of the partition. 
By maximising the resulting value, we are minimising the number of missing coverages (Eq. \eqref{eq:miss_cov}).

To compute the maximal $n$-soft domatic partition the objective function is adapted as follows:
\begin{equation}\label{eq:second_lp}
	\begin{array}{ll}
		\max & \sum_{v\in V} f\left(n^{-1}\cdot\sum_{i=1}^nf\left(\sum_{w\in N[v]} x_i^w\right)\right)
	\end{array}
\end{equation}
The term $n^{-1}\cdot\sum_{i=1}^nf\left(\sum_{w\in N[v]} x_i^w\right)$ describes the portion of sets of the partition having at least one common member with the set $N[v]$. 
For the maximal $n$-soft domatic partition it only matters whether a node's neighbourhood $N[v]$ has common members with all sets of the partition. 
Hence, we map the result to $0$ or $1$ and maximise the sum of those values. 
The LP applying this objective function minimises the number of incompletely covered nodes (Eq. \eqref{eq:inc_nodes}) by maximising the number of fully covered nodes.

Linear solvers are not able to solve objective functions with case distinctions directly. 
So, it is necessary to replace them. 
Therefore, we reformulate the LP to fit the standard form introduced in Equation \eqref{eq:sf_lp}. 

To do so, we introduce a set of auxiliary variables and additional constraints: 
\begin{align}\label{eq:optimal_domatic}
	\begin{array}{lll}
		\max        & \sum_{i=1}^n \sum_{v\in V} y_i^v\\
		\text{s.t.} & \forall v\in V:                              & \sum_{i=1}^n x_i^v=1\\
					& \forall v\in V,\;\forall i\in\{1,\ldots,n\}: & y_i^v\leq\sum_{w\in N[v]}x_i^w\\
					& \forall v\in V,\;\forall i\in\{1,\ldots,n\}: & x_i^v, y_i^v\in\{0,1\}
	\end{array}
\end{align}
The first new set of constraints $\forall v\in V,\;\forall i\in\{1,\ldots,n\}: y_i^v\leq\sum_{w\in N[v]}x_i^w$ ensures that the auxiliary variable $y_i^v$ is set to $1$ if in $N[v]$ exists a node included in set $i$ of the partition.
Therefore, if there are multiple nodes of $N[v]$ in the set $i$ of the partition it does not affect the outcome of our LP because $y_i^v$ is a binary variable and cannot grow larger than $1$. 
The resulting objective function maximises the sum of all $y_i^v$. 
Therefore, it replaces our auxiliary function $f$.

For the maximal $n$-soft domatic partition, we repeat the pattern applied to Equation \eqref{eq:optimal_domatic} in similar fashion: 
\begin{align}\label{eq:maximal_domatic}
	\begin{array}{lll}
		\max        & \sum_{v\in V} z^v\\
		\text{s.t.} & \forall v\in V:                              & \sum_{i=1}^n x_i^v=1\\
					& \forall v\in V,\;\forall i\in\{1,\ldots,n\}: & y_i^v\leq\sum_{w\in N[v]}x_i^w\\
					& \forall v\in V,\;\forall i\in\{1,\ldots,n\}: & z^v\leq y_i^v\\
					& \forall v\in V,\;\forall i\in\{1,\ldots,n\}: & x_i^v, y_i^v\in\{0,1\}
	\end{array}
\end{align}
Rather than aggregating the $y_i^v$ as the number of sets of the partition that intersect non-empty with $N[v]$, we establish $z_v=1$ under the condition that their are no empty intersections. 
Again, $y_i^v$ is $1$ if $N[v]$ incorporates at least one node of the set $i$ of the partition.
Additionally, we introduce the set of auxiliary variables $z^v\in \{0,1\}$. 
The constraint $\forall v\in V,\;\forall i\in \{1,\ldots,n\}: z^v\leq y_i^v$ and objective function ensure $z^v$ is equal to the largest $y_i^v$. 
Hence, the LP minimise the number of incompletely covered nodes from Equation \eqref{eq:inc_nodes} by maximising the number of completely covered nodes.

The LP for the optimal as well as the maximal $n$-soft domatic partition can also be modified to minimise the number of missing coverages or incompletely covered nodes if a node incorporates more than one security mean. 
We have discussed two versions of this approach: 
Either by implementing a fix number of security means per node or by distributing different combinations of security means based on their individual estimated costs. 
If a node is allowed to implement a fix number of $k\in \mathbb{N}_{>1}$ different security means, the constraint $\forall v\in V:\sum_{i=1}^n x_i^v=1$ changes to: 
\begin{equation}\label{eq:sec_means_k}
	\forall v\in V: \sum_{i=1}^n x_i^v=k
\end{equation}

Next, we apply security means (associated with sets $i$ of the partition) based on the share of resources necessary per security mean $m_i$, available at each node $v\in V$.
The constraint $\forall v\in V:\sum_{i=1}^n x_i^v=1$ has to be updated as follows:
\begin{equation}\label{eq:sec_means_perf}
	\forall v\in V: \sum_{i=1}^n m_i \cdot x_i^v=1
\end{equation}
The resource costs over all security means form a vector $\mathbf{m} \in \mathbb{R}^n$ with its components $m_i\in (0,1]$.
Without loss of generality, the overall resources available for security means per node have been set to $1$.
Each value $m_i$ represents the individual portion of costs caused for operating security mean $i$ in relation to the total costs for all security means.
The representation of necessary and available resources as a scalar is a simplification showing the feasibility of our LPs to take those into account.

\section{$\lambda$-precision UDG Generator}\label{sec:udg_gen}
The algorithms we propose to distribute security means in favour of the neighbourhood watch inspired security framework for large-scale static homogeneous WSNs are NP hard. 
It is necessary to validate the computability of the algorithms on a large number of realistic WSN models. 
Since, we cannot pinpoint the exact influence of graph properties on the computation time of our partitioning algorithms, we examine it empirically. 
Computations on a large set of models enable us to study the relation between different graph properties and the computation time.
To do so, we need a generator supplying it with a large variety and number of WSN graph models with desired properties. 

With the growing demand and sizes of WSNs \cite{el2022wireless}, the attention of potential attackers \cite{pawar2017literature} increases as well.
As consequence, more complex security \cite{kumar2020analysis} and communication protocols \cite{ketshabetswe2019communication} are developed. 
The application of those protocols leads to an increasing power consumption which affects the available computational and energy resources for the actual tasks of nodes. 
Since nodes and their distribution are expensive and their failure can lead to the failure of the network, network operators are interested in maximising the potential lifetime of nodes and the networks. 
An attempt to deal with the higher demand in power are smart sleep scheduling schemes \cite{kovasznai2018investigations} and hop-by-hop communication strategies \cite{basaran2010hop}. 
Additionally, there are many algorithms whose complexities exceed the deterministic polynomial time bound or are bound to higher polynomial degrees \cite{ding2008adaptive,saha2014distributed}. 
For researchers to decide whether those algorithms can be solved analytically or bring the need of an approximation, empirical evaluations on WSN graph models with desired properties are necessary.
To generate these models, we introduce a graph generator that creates $\lambda$-precision UDGs by distributing nodes randomly and uniformly in a unit square.

$\lambda$-precision UDGs have several advantages compared to ordinary UDGs. 
The $\lambda$-precision limits the node degree of each node in the graph. 
The limitation results from the size of the ring given by the radii $\lambda$ and $r_\text{tr}$ with $0<\lambda< r_\text{tr}$. 
Nodes only connect (have common edges) to nodes within this ring, since each node has to have at least $\lambda$ distance to other nodes in the network. 
Therefore, there is an upper limit of nodes that are able to connect. 
Choosing the $\lambda$ distance and node number $|V|$ so that a large portion of the area of the generation is covered ensures that the nodes are more evenly spaced out on the generation plane.
Hence, it allows to control the variance of the local cluster coefficient.
With regard to WSNs, evenly distributed nodes improve the area-wide monitoring.

\subsection{Node Distribution}\label{sec:node_distribution}
To generate the graphs, we start with randomly and uniformly distributing nodes in a unit square with the constraint that two nodes have to have a minimal distance $\lambda$ in between them. 
For an efficient computation, it is necessary to discretise the unit square. 
We do so, with a uniform grid size of $1000$ times $1000$. 
The grid size can be adapted as needed and is often chosen based on the computational limits and the intended graph properties as for example the number of nodes.
In the implementation, we distribute the nodes iteratively.
Each node occupies the grid coordinate of its centre and all grid coordinates within $\lambda$ distance from it.
For this purpose, each grid coordinate gets assigned a marker value. 
The marker indicates whether the coordinate is still {\itshape available} ($0$) or {\itshape occupied} ($1$). 
After a new node has been added, all surrounding marker values in $\lambda$ distance are updated by setting them to $1$.
The coordinate for the centre of the succeeding nodes is randomly and uniformly selected from the non-occupied coordinates. 
The process is repeated until either no grid coordinates are available or the desired amount of nodes has been placed within the unit square.

\subsection{Generator Seeds}\label{sec:gen_seeds} 
\begin{table}[bh!]
	\centering
	\caption{Empirically determined seeds to generate graphs with an expected average node degree in between $\deg_\text{exp}$ to $\deg_\text{exp} + 0.25$ for a given number of nodes $|V|$ and a desired medium total coverage of the generation plane $\overline{A_\text{coverage}}$ from $85\%$ to $87.5\%$. The values have been determined by generating repeatedly sets of $20$ graphs for varying values of $\lambda$ and $r_\text{tr}$ until approaching the desired properties. The probability $P_\text{connected}$ is the empirically determined likeliness of a graph to be connected for the given parameters. The results for $\overline{A_\text{coverage}}$, $\overline{\deg_\text{avg}}$ and $P_\text{connected}$ are the arithmetic mean values of $20$ graphs of the determined input parameter combinations.}
	\label{tbl:test}
	\scriptsize
	\rowcolors{2}{gray!25}{white}
	\begin{tabular}{rr|rr|rrr}
		\specialrule{.5pt}{0pt}{.5pt}
		\raggedleft $|V|$ & 
		\raggedleft $\deg_\text{exp}$ & 
		\raggedleft $\lambda$ & 
		\raggedleft $r_\text{tr}$ & 
		\raggedleft $\overline{A_\text{coverage}}$ & 
		\raggedleft $\overline{\deg_\text{avg}}$ &
		\raggedleft $P_\text{connected}$ \tabularnewline
		\specialrule{.5pt}{.5pt}{.5pt}
        $20 $ & $3$ & $0.148$ & $0.290$ & $0.855$ & $3.105$ & $0.75$ \\
        $20 $ & $4$ & $0.148$ & $0.333$ & $0.855$ & $4.005$ & $0.95$ \\
        $20 $ & $5$ & $0.148$ & $0.383$ & $0.855$ & $5.210$ & $1.00$ \\
        $20 $ & $6$ & $0.148$ & $0.411$ & $0.855$ & $6.055$ & $1.00$ \\
        $40 $ & $3$ & $0.104$ & $0.196$ & $0.869$ & $3.055$ & $0.50$ \\
        $40 $ & $4$ & $0.104$ & $0.226$ & $0.869$ & $4.140$ & $0.85$ \\
        $40 $ & $5$ & $0.104$ & $0.250$ & $0.869$ & $5.100$ & $1.00$ \\
        $40 $ & $6$ & $0.104$ & $0.277$ & $0.869$ & $6.195$ & $1.00$ \\
        $60 $ & $3$ & $0.085$ & $0.159$ & $0.862$ & $3.198$ & $0.05$ \\
        $60 $ & $4$ & $0.085$ & $0.179$ & $0.862$ & $4.156$ & $0.70$ \\
        $60 $ & $5$ & $0.085$ & $0.197$ & $0.862$ & $5.075$ & $0.95$ \\
        $60 $ & $6$ & $0.085$ & $0.219$ & $0.862$ & $6.198$ & $0.95$ \\
        $80 $ & $3$ & $0.072$ & $0.136$ & $0.854$ & $3.222$ & $0.20$ \\
        $80 $ & $4$ & $0.072$ & $0.152$ & $0.854$ & $4.061$ & $0.65$ \\
        $80 $ & $5$ & $0.072$ & $0.168$ & $0.854$ & $5.067$ & $0.80$ \\
        $80 $ & $6$ & $0.072$ & $0.181$ & $0.854$ & $6.008$ & $0.90$ \\
        $100$ & $3$ & $0.065$ & $0.120$ & $0.861$ & $3.142$ & $0.20$ \\
        $100$ & $4$ & $0.065$ & $0.137$ & $0.861$ & $4.234$ & $0.70$ \\
        $100$ & $5$ & $0.065$ & $0.150$ & $0.861$ & $5.127$ & $0.85$ \\
        $100$ & $6$ & $0.065$ & $0.164$ & $0.861$ & $6.198$ & $1.00$ \\
        $120$ & $3$ & $0.059$ & $0.108$ & $0.855$ & $3.120$ & $0.10$ \\
        $120$ & $4$ & $0.059$ & $0.122$ & $0.855$ & $4.086$ & $0.70$ \\
        $120$ & $5$ & $0.059$ & $0.135$ & $0.855$ & $5.115$ & $1.00$ \\
        $120$ & $6$ & $0.059$ & $0.147$ & $0.855$ & $6.196$ & $1.00$ \\
        $140$ & $3$ & $0.054$ & $0.098$ & $0.866$ & $3.074$ & $0.10$ \\
        $140$ & $4$ & $0.054$ & $0.112$ & $0.866$ & $4.127$ & $0.70$ \\
        $140$ & $5$ & $0.054$ & $0.124$ & $0.866$ & $5.089$ & $0.95$ \\
        $140$ & $6$ & $0.054$ & $0.136$ & $0.866$ & $6.237$ & $1.00$ \\
        $160$ & $3$ & $0.051$ & $0.094$ & $0.857$ & $3.215$ & $0.05$ \\
        $160$ & $4$ & $0.051$ & $0.105$ & $0.857$ & $4.057$ & $0.60$ \\
        $160$ & $5$ & $0.051$ & $0.116$ & $0.857$ & $5.164$ & $0.90$ \\
        $160$ & $6$ & $0.051$ & $0.127$ & $0.857$ & $6.183$ & $1.00$ \\
        $180$ & $3$ & $0.048$ & $0.087$ & $0.853$ & $3.093$ & $0.00$ \\
        $180$ & $4$ & $0.048$ & $0.098$ & $0.853$ & $4.063$ & $0.50$ \\
        $180$ & $5$ & $0.048$ & $0.109$ & $0.853$ & $5.128$ & $0.90$ \\
        $180$ & $6$ & $0.048$ & $0.117$ & $0.853$ & $6.064$ & $0.95$ \\
        $200$ & $3$ & $0.045$ & $0.082$ & $0.866$ & $3.095$ & $0.00$ \\
        $200$ & $4$ & $0.045$ & $0.093$ & $0.866$ & $4.139$ & $0.55$ \\
        $200$ & $5$ & $0.045$ & $0.102$ & $0.866$ & $5.047$ & $0.75$ \\
        $200$ & $6$ & $0.045$ & $0.111$ & $0.866$ & $6.056$ & $0.95$ \\
        $220$ & $3$ & $0.044$ & $0.077$ & $0.867$ & $3.031$ & $0.05$ \\
        $220$ & $4$ & $0.044$ & $0.089$ & $0.867$ & $4.164$ & $0.65$ \\
        $220$ & $5$ & $0.044$ & $0.098$ & $0.867$ & $5.132$ & $0.90$ \\
        $220$ & $6$ & $0.044$ & $0.107$ & $0.867$ & $6.179$ & $1.00$ \\
        $240$ & $3$ & $0.041$ & $0.075$ & $0.863$ & $3.138$ & $0.05$ \\
        $240$ & $4$ & $0.041$ & $0.084$ & $0.863$ & $4.011$ & $0.25$ \\
        $240$ & $5$ & $0.041$ & $0.093$ & $0.863$ & $5.041$ & $0.75$ \\
        $240$ & $6$ & $0.041$ & $0.101$ & $0.863$ & $6.047$ & $0.85$ \\
        $260$ & $3$ & $0.040$ & $0.070$ & $0.855$ & $3.006$ & $0.00$ \\
        $260$ & $4$ & $0.040$ & $0.081$ & $0.855$ & $4.116$ & $0.30$ \\
        $260$ & $5$ & $0.040$ & $0.089$ & $0.855$ & $5.088$ & $0.65$ \\
        $260$ & $6$ & $0.040$ & $0.097$ & $0.855$ & $6.167$ & $0.95$ \\
        $280$ & $3$ & $0.039$ & $0.067$ & $0.867$ & $3.025$ & $0.00$ \\
        $280$ & $4$ & $0.039$ & $0.077$ & $0.867$ & $4.080$ & $0.60$ \\
        $280$ & $5$ & $0.039$ & $0.086$ & $0.867$ & $5.082$ & $0.85$ \\
        $280$ & $6$ & $0.039$ & $0.094$ & $0.867$ & $6.183$ & $0.95$ \\
        $300$ & $3$ & $0.037$ & $0.066$ & $0.874$ & $3.008$ & $0.05$ \\
        $300$ & $4$ & $0.037$ & $0.074$ & $0.874$ & $4.036$ & $0.60$ \\
        $300$ & $5$ & $0.037$ & $0.083$ & $0.874$ & $5.029$ & $0.95$ \\
        $300$ & $6$ & $0.037$ & $0.090$ & $0.874$ & $6.014$ & $1.00$ \\
		\specialrule{.5pt}{.0pt}{0pt}
	\end{tabular}
\end{table}

In order to create $\lambda$-precision UDGs with specific properties, it is essential to determine the input parameters (generator seeds) resulting in graphs with the desired properties.
We use the following input parameters: number of nodes $|V|$, the pairwise minimal distance in between the nodes $\lambda$ and the distance $r_\text{tr}$ up to which nodes are connected.
For the empirical evaluation of random graphs, we compute for each parameter set a certain amount of graphs. 
After computing a set of graphs for chosen input parameters, we compare the properties with our target values. 
Depending on the outcome, we either save the result or adjust the input parameters.
To do so, we apply a binary search separately for both parameters $\lambda$ and $r_\text{tr}$, starting with $\lambda$.
Table \ref{tbl:test} shows the resulting generator seeds to create random $\lambda$-precision UDGs with desired properties. 
As shown in Table \ref{tbl:test} the target values are the medium total coverage of the generation plane $\overline{A_\text{coverage}}$ and the medium average node degree $\overline{\deg_\text{avg}}$.

The value of $\overline{A_\text{coverage}}$ affects the probability of resulting random $\lambda$-precision UDGs to be connected $P_\text{connected}$. 
In addition, it ensures a low variance of the local cluster coefficient and an even coverage of the generation plane as we discuss in Subsection \ref{subsec:cluster_coefficient}. 
Applying a binary search, we first approach the radius $\lambda$ achieving a medium total coverage of the generation plane $\overline{A_\text{coverage}}$ between $85\%$ and $87.5\%$. 
The coverage is determined numerically. 
After distributing nodes as described in the previous subsection, the relation between occupied grid coordinates and the grid size yields the total coverage of the generation plane $A_\text{coverage}$. 
Finally, the medium total coverage of the generation plane $\overline{A_\text{coverage}}$ of all generated graphs for the input parameter set is computed.

Next, we determine the transmission range $r_\text{tr}$ using a binary search until we reach a medium average node degree $\overline{\deg_\text{avg}}$ over all graphs obtained for the given input parameters. 
The final results of the computed generator seeds is shown in Table \ref{tbl:test}. 
To generate the graphs for the evaluation of our $0-1$ LPs we will use the results from this table.

\subsection{Cluster Coefficient and Degree Distribution}\label{subsec:cluster_coefficient}
We show that the variance of the local cluster coefficient and the variance of the node degree distribution decreases with an increasing coverage of the generation plane.
This allows to generate specific graphs and test the effect of those properties on the computation time of our partitions.
We can assume that a graph contains an edge, or a small number of edges, whose removal would disconnect the graph into several larger connected components. 
In such cases, we can expect a decrease in computation time compared to a graph without such breaking points.
Therefore, we expect that a low variance of the distribution of the node degrees provides information about an upper bound of the computation time.
The portion of the covered area of the generation plane is directly linked to the combined choice of the number of nodes $|V|$ and their minimal pairwise distance $\lambda$ within the graph. 
To evaluate the behaviour of the interplay between the portion of the covered area of the generation plane and the variance of the node degree distribution as well as the variance of the local cluster coefficient we determine additional generator seeds.
To do so, we first determine generator seeds for selected target values as in the previous section. 
We have chosen an expected average node degree $\deg_\text{exp}$ of $4$ and $5$. 
The observed node numbers $|V|$ are $100$ and $200$. 
For each of those combinations the expected covered area of the generation plane is set to the intervals $\{[0.850, 0.875]$, $[0.875,0.900]$, $[0.900,0.925]$, $[0.925,0.950]$, $[0.950,0.975]$, $[0.975,1.000]\}$.
We exhibit the determined generator seeds in Table \ref{tbl:local_clustering}.
\begin{table}[b!]
	\scriptsize
	\caption{The seeds for our generator to test the behaviour of the variance of the local cluster coefficient and the variance of the node degree distribution subject to the total coverage of the generation plane have been computed as in Table \ref{tbl:test}. 
		The abbreviation $A_\text{exp\_cov}$ stands for the expected coverage area. 
		It represents the coverage range for which we determine $\lambda$ and $r_\text{tr}$. 
		Therefore, the resulting coverage area $A_\text{cov}$ of graphs generated with those parameters is likely to be within the specified range. 
	$\overline{A_\text{cov}}$ is short for $\overline{A_\text{coverage}}$ and $P_\text{conn}$ abbreviates $P_\text{connected}$ according to Table \ref{tbl:test}.}
	\label{tbl:local_clustering}
	\rowcolors{2}{gray!25}{white}
	\begin{tabular}{rrr|rr|rrr}
		\specialrule{.5pt}{0pt}{.5pt}
		\raggedleft $A_\text{exp\_cov}$ & 
		\raggedleft $|V|$ & 
		\raggedleft $\deg_\text{exp}$ & 
		\raggedleft $\lambda$ & 
		\raggedleft $r_\text{tr}$ & 
		\raggedleft $\overline{A_\text{cov}}$ & 
		\raggedleft $\overline{\deg_\text{avg}}$ &
		\raggedleft $P_\text{conn}$ \tabularnewline
		\specialrule{.5pt}{.5pt}{.5pt}
		$[0.850, 0.875]$ & $100$ & $4$ & $0.065$ & $0.136$ & $0.857$ & $4.193$ & $0.575$ \\
		$[0.850, 0.875]$ & $100$ & $5$ & $0.065$ & $0.148$ & $0.857$ & $5.073$ & $0.825$ \\
		$[0.850, 0.875]$ & $200$ & $4$ & $0.045$ & $0.094$ & $0.864$ & $4.174$ & $0.525$ \\
		$[0.850, 0.875]$ & $200$ & $5$ & $0.045$ & $0.103$ & $0.864$ & $5.093$ & $0.850$ \\
		$[0.875, 0.900]$ & $100$ & $5$ & $0.066$ & $0.148$ & $0.890$ & $5.044$ & $0.975$ \\
		$[0.875, 0.900]$ & $100$ & $4$ & $0.066$ & $0.136$ & $0.890$ & $4.161$ & $0.800$ \\
		$[0.875, 0.900]$ & $200$ & $4$ & $0.047$ & $0.094$ & $0.884$ & $4.163$ & $0.650$ \\
		$[0.875, 0.900]$ & $200$ & $5$ & $0.047$ & $0.103$ & $0.884$ & $5.070$ & $0.900$ \\
		$[0.900, 0.925]$ & $100$ & $4$ & $0.068$ & $0.136$ & $0.915$ & $4.143$ & $0.800$ \\
		$[0.900, 0.925]$ & $100$ & $5$ & $0.068$ & $0.152$ & $0.915$ & $5.249$ & $0.925$ \\
		$[0.900, 0.925]$ & $200$ & $4$ & $0.048$ & $0.094$ & $0.918$ & $4.125$ & $0.600$ \\
		$[0.900, 0.925]$ & $200$ & $5$ & $0.048$ & $0.105$ & $0.918$ & $5.224$ & $0.900$ \\
		$[0.925, 0.950]$ & $100$ & $4$ & $0.070$ & $0.136$ & $0.932$ & $4.094$ & $0.950$ \\
		$[0.925, 0.950]$ & $100$ & $5$ & $0.070$ & $0.153$ & $0.932$ & $5.231$ & $1.000$ \\
		$[0.925, 0.950]$ & $200$ & $5$ & $0.051$ & $0.103$ & $0.947$ & $5.028$ & $0.975$ \\
		$[0.925, 0.950]$ & $200$ & $4$ & $0.051$ & $0.094$ & $0.947$ & $4.109$ & $0.700$ \\
		$[0.950, 0.975]$ & $100$ & $5$ & $0.074$ & $0.153$ & $0.968$ & $5.179$ & $1.000$ \\
		$[0.950, 0.975]$ & $100$ & $4$ & $0.074$ & $0.136$ & $0.968$ & $4.083$ & $0.900$ \\
		$[0.950, 0.975]$ & $200$ & $5$ & $0.051$ & $0.103$ & $0.959$ & $5.006$ & $1.000$ \\
		$[0.950, 0.975]$ & $200$ & $4$ & $0.051$ & $0.094$ & $0.959$ & $4.114$ & $0.875$ \\
		$[0.975, 1.000]$ & $100$ & $4$ & $0.078$ & $0.136$ & $0.987$ & $4.114$ & $0.950$ \\
		$[0.975, 1.000]$ & $100$ & $5$ & $0.078$ & $0.153$ & $0.987$ & $5.104$ & $1.000$ \\
		$[0.975, 1.000]$ & $200$ & $4$ & $0.059$ & $0.094$ & $0.999$ & $4.157$ & $0.900$ \\
		$[0.975, 1.000]$ & $200$ & $5$ & $0.059$ & $0.105$ & $0.999$ & $5.074$ & $1.000$ \\
		\specialrule{.5pt}{.0pt}{0pt}
	\end{tabular}
\end{table}
The results indicate that the determined seeds maintain a high probability to create connected graphs even with a decreasing medium total coverage of the generation plane $\overline{A_\text{coverage}}$.

\begin{figure}[t]
	\begin{subfigure}{\linewidth}
		\includegraphics[width=.24\linewidth]{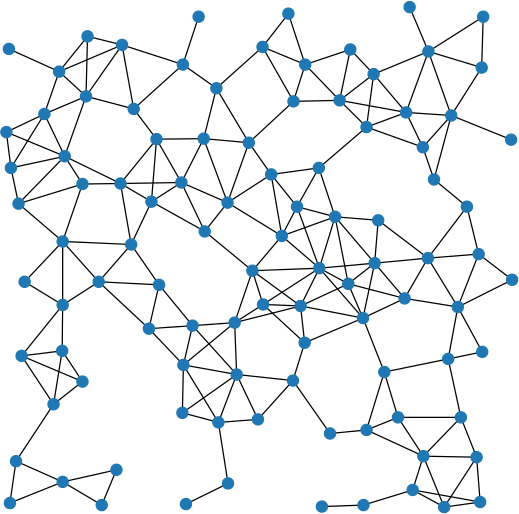}\hfill
		\includegraphics[width=.24\linewidth]{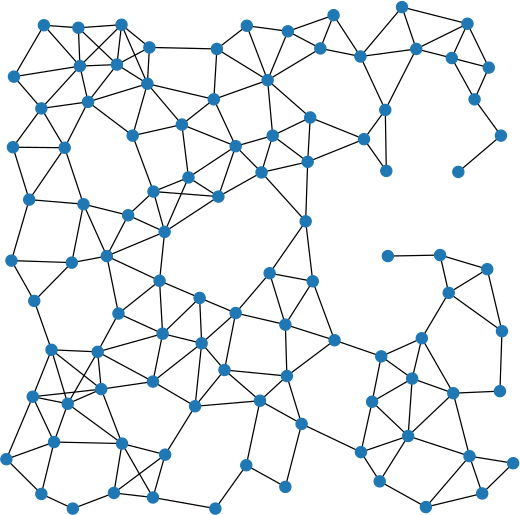}\hfill
		\includegraphics[width=.24\linewidth]{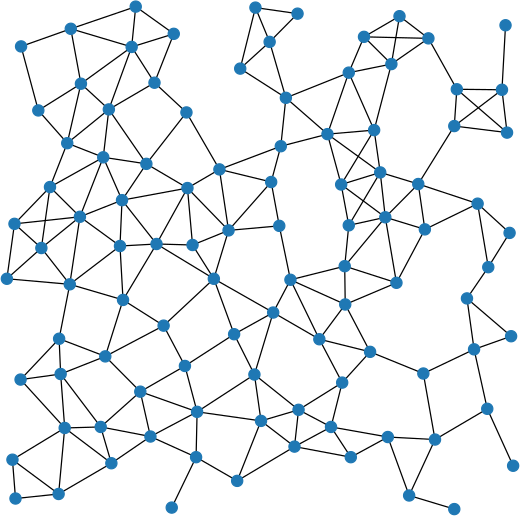}\hfill
		\includegraphics[width=.24\linewidth]{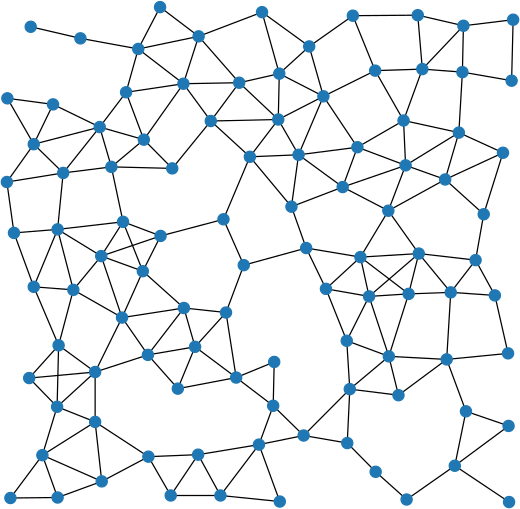}
		\captionsetup{justification=centering}
		\caption{$|V| = 100$, $\deg_\text{avg}=4.0$,\\[2pt] $\lambda$ (left to right): $0.065$, $0.068$, $0.070$, $0.078$}
	\end{subfigure}\par\medskip
	\begin{subfigure}{\linewidth}
		\includegraphics[width=.24\linewidth]{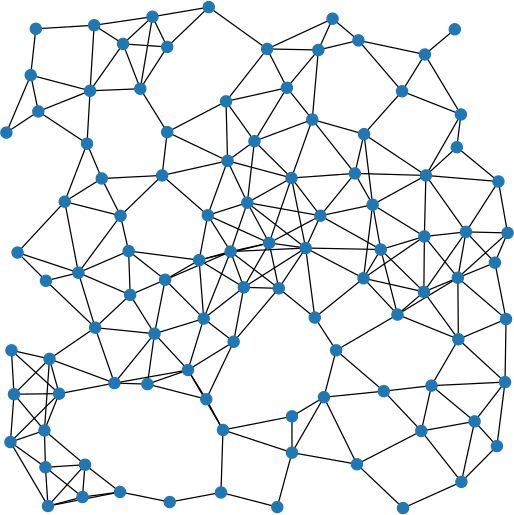}\hfill
		\includegraphics[width=.24\linewidth]{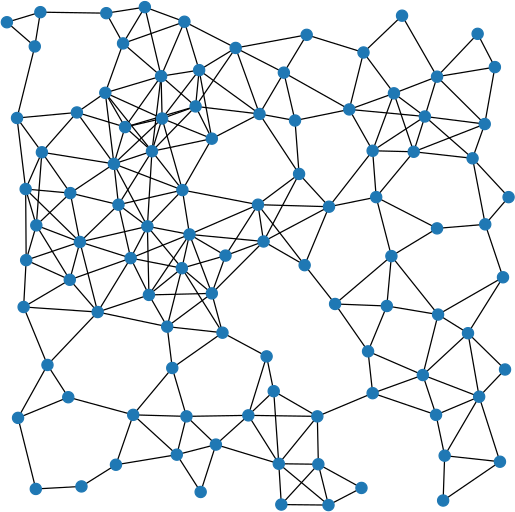}\hfill
		\includegraphics[width=.24\linewidth]{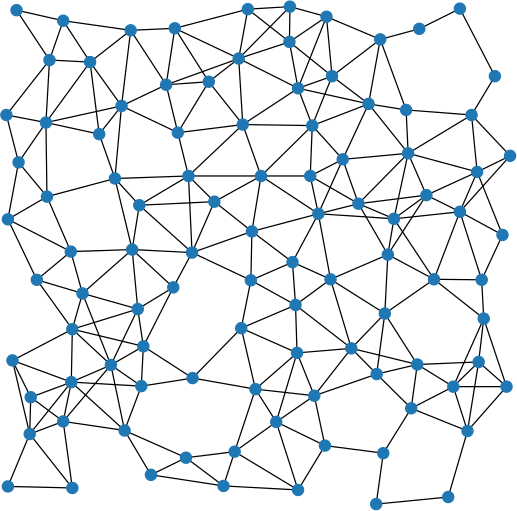}\hfill
		\includegraphics[width=.24\linewidth]{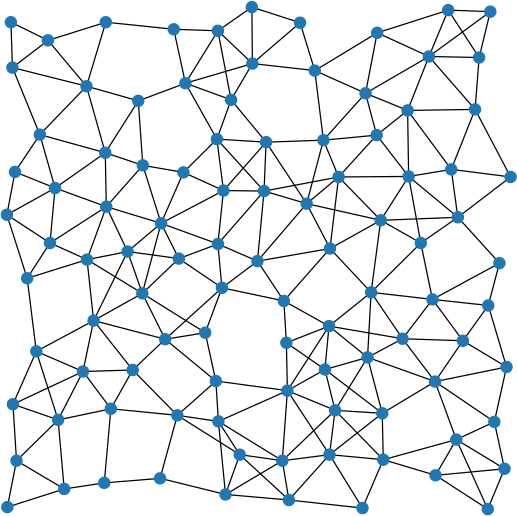}
		\captionsetup{justification=centering}
		\caption{$|V| = 100$, $\deg_\text{avg}=5.0$,\\[2pt] $\lambda$ (left to right): $0.065$, $0.068$, $0.070$, $0.078$}
	\end{subfigure}\par\medskip
	\begin{subfigure}{\linewidth}
		\includegraphics[width=.24\linewidth]{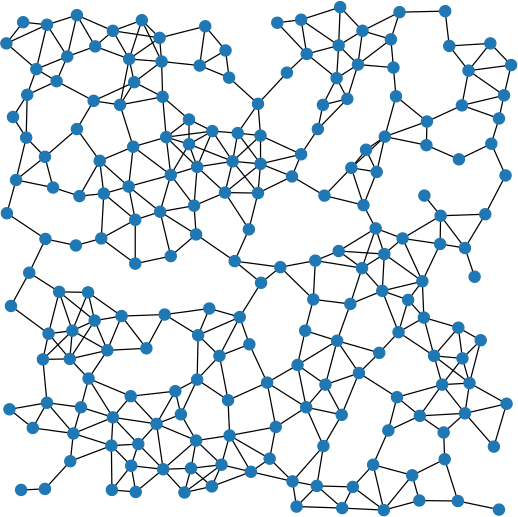}\hfill
		\includegraphics[width=.24\linewidth]{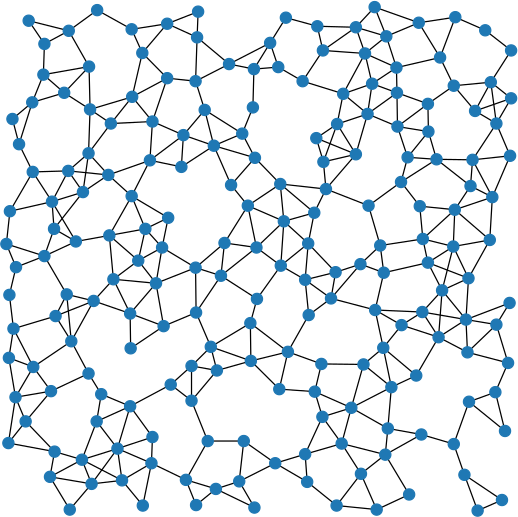}\hfill
		\includegraphics[width=.24\linewidth]{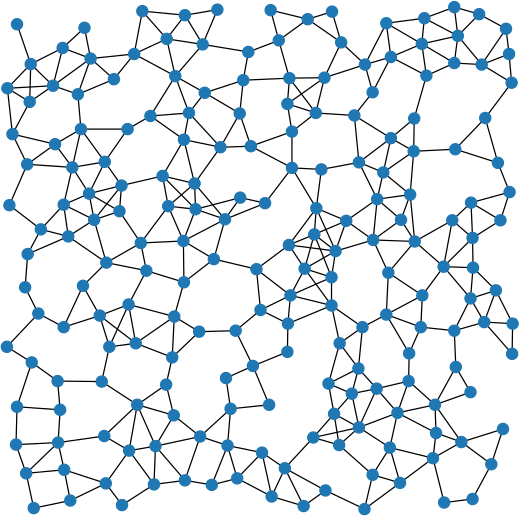}\hfill
		\includegraphics[width=.24\linewidth]{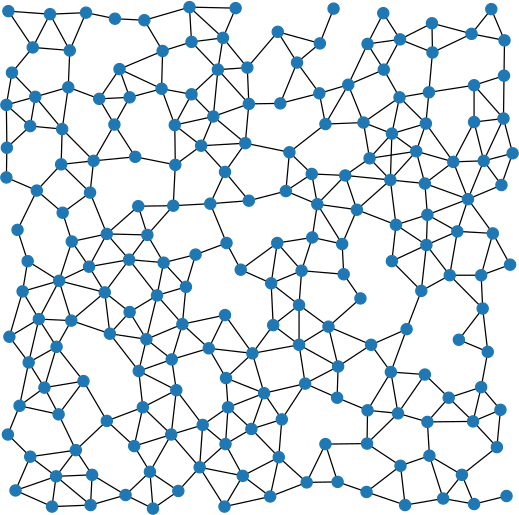}
		\captionsetup{justification=centering}
		\caption{$|V| = 200$, $\deg_\text{avg}=4.0$,\\[2pt] $\lambda$ (left to right): $0.045$, $0.048$, $0.051$, $0.059$}
	\end{subfigure}\par\medskip
	\begin{subfigure}{\linewidth}
		\includegraphics[width=.24\linewidth]{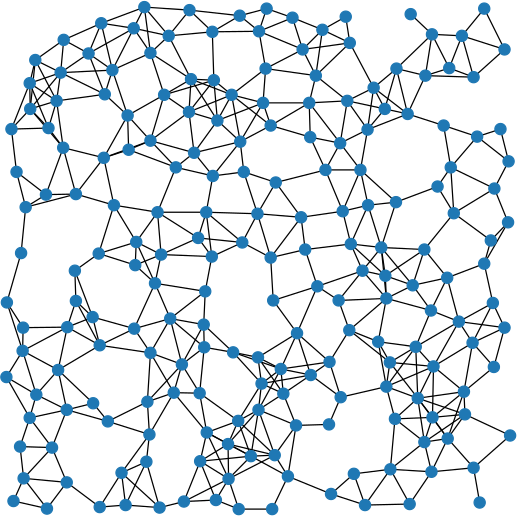}\hfill
		\includegraphics[width=.24\linewidth]{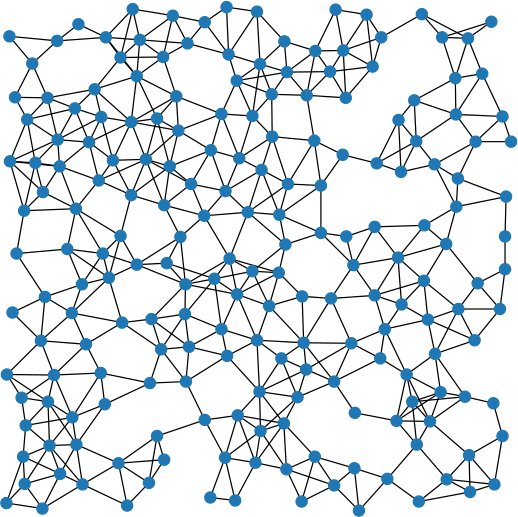}\hfill
		\includegraphics[width=.24\linewidth]{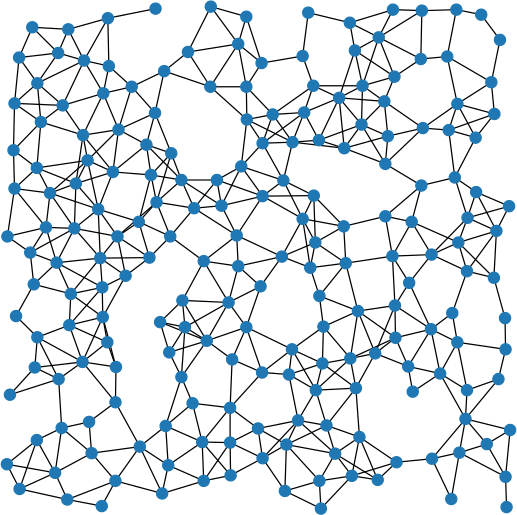}\hfill
		\includegraphics[width=.24\linewidth]{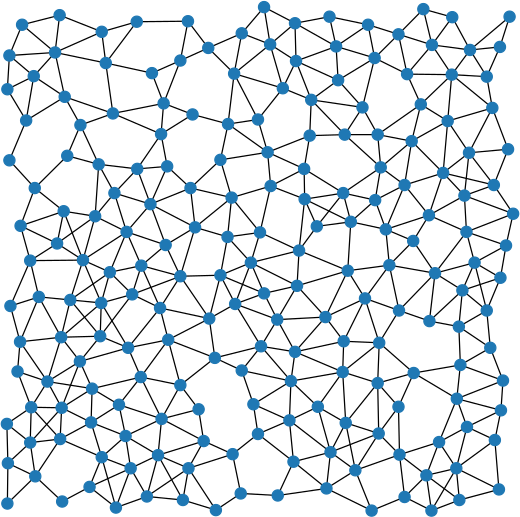}
		\captionsetup{justification=centering}
		\caption{$|V| = 200$, $\deg_\text{avg}=5.0$,\\[2pt] $\lambda$ (left to right): $0.045$, $0.048$, $0.051$, $0.059$}
	\end{subfigure}
	\caption{As example for the resulting $\lambda$-precision UDGs shown per row are from left to right generated for $A_\text{coverage}$ of the ranges $[0.850,0.875]$, $[0.900,0.925]$, $[0.925,0.950]$ and $[0.975,1.000]$ respectively.}
	\label{fig:example_graphs_area_coverage}
\end{figure}

For our empirical analysis of the relation between the medium total coverage of the generation plane $\overline{A_\text{coverage}}$ and the variance of the node degree distribution as well as the variance of the local cluster coefficient, we determine the parameters with the same binary search utilised in the previous subsection. 
By means of these parameters, we compute $40$ sample graphs for each of the discussed target parameter combinations. 
The target parameters are number of nodes $|V|$, medium average node degree $\overline{\deg_\text{avg}}$ and medium total coverage of the generation plane $\overline{A_\text{coverage}}$. 
A selection of $16$ of the resulting uniformly and randomly determined $\lambda$-precision UDGs is displayed in Fig. \ref{fig:example_graphs_area_coverage}.

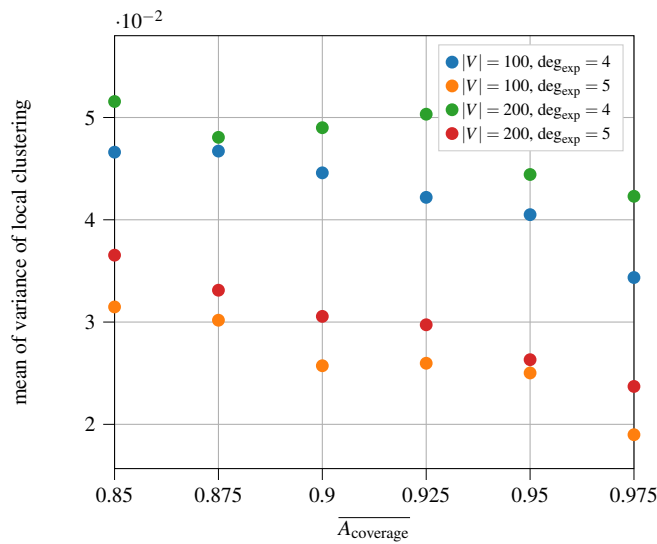
\begin{figure}[t]
	\centering
\begin{tikzpicture}

\definecolor{crimson2143940}{RGB}{214,39,40}
\definecolor{darkgray176}{RGB}{176,176,176}
\definecolor{darkorange25512714}{RGB}{255,127,14}
\definecolor{forestgreen4416044}{RGB}{44,160,44}
\definecolor{gainsboro229}{RGB}{229,229,229}
\definecolor{lightgray204}{RGB}{204,204,204}
\definecolor{steelblue31119180}{RGB}{31,119,180}

\definecolor{blue023255}{RGB}{0,23,255}
\definecolor{darkgray176}{RGB}{176,176,176}
\definecolor{gainsboro229}{RGB}{229,229,229}
\definecolor{lightgray204}{RGB}{204,204,204}
\definecolor{red255040}{RGB}{255,0,40}
\definecolor{springgreen0255139}{RGB}{0,255,139}
\definecolor{yellow2052550}{RGB}{205,255,0}

\begin{axis}[
legend cell align={left},
legend style={
	fill opacity=1, 
	draw opacity=1, 
	text opacity=1, 
	draw=lightgray204,
	nodes={scale=0.8, transform shape}
},
tick align=outside,
tick pos=left,
x grid style={darkgray176},
xlabel={\(\displaystyle \overline{A_\text{coverage}}\)},
xmajorgrids,
xmin=0.85, xmax=0.975,
xtick style={color=black},
xtick={0.85, 0.875, 0.9, 0.925, 0.95, 0.975}, 
xticklabels={0.85, 0.875, 0.9, 0.925, 0.95, 0.975}, 
y grid style={darkgray176},
ylabel={mean of variance of local clustering},
ymajorgrids,
ymin=0.0156726435090703, ymax=0.058,
ytick style={color=black},
style={/pgf/number format/fixed}
]
\addplot [thick, steelblue31119180, only marks, mark=*]
table {%
0.85 0.0466104280655549
0.875 0.0467174120559335
0.9 0.0445900192247733
0.925 0.0421934519982993
0.95 0.0405012611181973
0.975 0.0343544135345805
};
\addlegendentry{$|V|=100$, $\deg_\text{exp}=4$}
\addplot [thick, darkorange25512714, only marks, mark=*]
table {%
0.85 0.0314798162548816
0.875 0.0301801024344923
0.9 0.0257298555472884
0.925 0.0259728155954743
0.95 0.0250242822837931
0.975 0.0189860680874433
};
\addlegendentry{$|V|=100$, $\deg_\text{exp}=5$}
\addplot [thick, forestgreen4416044, only marks, mark=*]
table {%
0.85 0.0515657275415722
0.875 0.0480611753495843
0.9 0.0490024711026077
0.925 0.0503231229591837
0.95 0.0444277697278912
0.975 0.0422969202947846
};
\addlegendentry{$|V|=200$, $\deg_\text{exp}=4$}
\addplot [thick, crimson2143940, only marks, mark=*]
table {%
0.85 0.0365403171166025
0.875 0.0331140850749559
0.9 0.0305559016250945
0.925 0.0297339729324137
0.95 0.0263204015353364
0.975 0.0237057566279289
};
\addlegendentry{$|V|=200$, $\deg_\text{exp}=5$}
\end{axis}

\end{tikzpicture}
	\caption{The mean of the variance of the local cluster coefficients tends to decrease along with increasing $\overline{A_\text{coverage}}$ leading to more homogeneously distributed nodes with larger pairwise distances. 
		A sample size of $40$ graphs per data point has been utilised. 
	This sample size balances the expressivity of the decrease trend with the computational effort for parameterised graph generation.}
	\label{fig:dia_local_clustering}
\end{figure}

\begin{figure}[t]
	\centering
\begin{tikzpicture}

\definecolor{crimson2143940}{RGB}{214,39,40}
\definecolor{darkorange25512714}{RGB}{255,127,14}
\definecolor{forestgreen4416044}{RGB}{44,160,44}
\definecolor{steelblue31119180}{RGB}{31,119,180}
\definecolor{darkgray176}{RGB}{176,176,176}

\definecolor{blue023255}{RGB}{0,23,255}
\definecolor{dimgray85}{RGB}{85,85,85}
\definecolor{gainsboro229}{RGB}{229,229,229}
\definecolor{lightgray204}{RGB}{204,204,204}
\definecolor{red255040}{RGB}{255,0,40}
\definecolor{springgreen0255139}{RGB}{0,255,139}
\definecolor{yellow2052550}{RGB}{205,255,0}

\begin{axis}[
axis line style={black},
legend cell align={left},
legend style={
	fill opacity=1.0, 
	draw opacity=1, 
	text opacity=1, 
	draw=lightgray204,
	nodes={scale=0.8, transform shape}
},
tick align=outside,
tick pos=left,
x grid style={darkgray176},
xlabel={\(\displaystyle \overline{A_\text{coverage}}\)},
xmajorgrids,
xmin=0.85, xmax=0.975,
xtick style={color=black},
xtick={0.85, 0.875, 0.9, 0.925, 0.95, 0.975}, 
xticklabels={0.85, 0.875, 0.9, 0.925, 0.95, 0.975}, 
y grid style={darkgray176},
ylabel={mean of variance of node degree distribution},
ymajorgrids,
ymin=1.16162925, ymax=2.4,
ytick style={color=black},
style={/pgf/number format/fixed}
]
\addplot [thick, steelblue31119180, only marks, mark=*]
table {%
0.85 1.835995
0.875 1.73878
0.9 1.6521575
0.925 1.5453625
0.95 1.47272
0.975 1.256155
};
\addlegendentry{$|V|=100$, $\deg_\text{exp}=4$}
\addplot [thick, darkorange25512714, only marks, mark=*]
table {%
0.85 2.346355
0.875 2.2652875
0.9 2.154345
0.925 1.8657475
0.95 1.762605
0.975 1.4520025
};
\addlegendentry{$|V|=100$, $\deg_\text{exp}=5$}
\addplot [thick, forestgreen4416044, only marks, mark=*]
table {%
0.85 2.0837
0.875 1.84541
0.9 1.7677
0.925 1.73226
0.95 1.54778
0.975 1.52709
};
\addlegendentry{$|V|=200$, $\deg_\text{exp}=4$}
\addplot [thick, crimson2143940, only marks, mark=*]
table {%
0.85 2.67383
0.875 2.46702
0.9 2.43598
0.925 2.4056
0.95 2.0809
0.975 1.93332
};
\addlegendentry{$|V|=200$, $\deg_\text{exp}=5$}
\end{axis}

\end{tikzpicture}
	\caption{The mean of the variance of the node degree distribution mostly diminishes subject to a growing $\overline{A_\text{coverage}}$. 
		This behaviour results in graphs that can be better employed for $n$-soft domatic partitions in a certain range for $n$. 
	Again, a sample size of $40$ graphs per data point has been utilised which implies some minor local fluctuations.}
	\label{fig:dia_node_deg_distribution}
\end{figure}
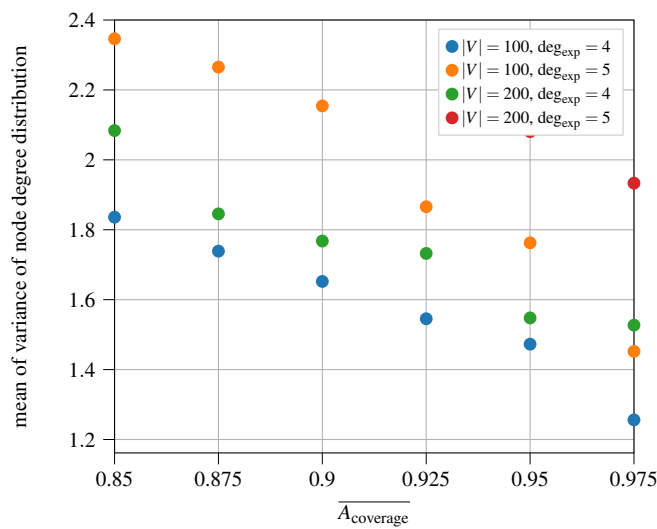

The results of our evaluation are depicted in Fig. \ref{fig:dia_local_clustering} and \ref{fig:dia_node_deg_distribution}. 
The $x$ coordinate for each data point is located at the lower value of the respective range representing the medium total coverage of the generation plane $\overline{A_\text{coverage}}$ in both diagrams. 
Each data point represents the arithmetic mean over the variance of the local cluster coefficients and the list of node degrees per graph for a sample size of $40$ graphs.

\subsection{Methods for Adaptation of Generated UDGs}\label{sec:graph_adapt}
Use case dependent adaptations of generated $\lambda$-precision UDGs can be necessary to satisfy certain requirements.
Therefore, to specify the accuracy of the graph properties, we have implemented methods to adapt the graphs resulting from our graph generator.
Connectivity, occurrences of bridges or average node degree are properties, we have considered for adaptation.
It is unlikely to receive a large randomly generated graph that meets exactly a set of desired properties based on selected input parameters.
To improve quality, validity and precision of an evaluation using graphs, it is desirable that those graphs meet exact criteria.
Certain properties are perhaps achievable solely by repeatedly generating graphs.
However, such a process is tedious and time consuming, especially for large numbers of graphs.\\
{\bfseries Connectivity:}
Our approach to connect a graph consisting of several connected components uses the nearest neighbour attempt. 
The algorithm determines all pairs of nearest neighbour nodes between distinct connected components. 
For each iteration, the nearest neighbour pair with the shortest edge length (euclidean distance) is chosen and added as edge to the graph.
Subsequently, all nearest neighbour pairs which in turn contain nodes from a single connected component inside the resulting graph are removed from the set. 
The last two steps are repeated until there is only one connected component in the graph.

Further, we assume that the occurrence of bridges in a graph model significantly affects the empirical test results and influences the evaluation of the computability of complex algorithms.
To validate this hypothesis, we provide an algorithm allowing to identify and to remove bridges. 
The result is a connected, bridge-free graph that serves as representation of large-scale static homogeneous WSNs. 
To identify possible bridges in our $\lambda$-precision UDGs, our generator utilises the NetworkX library. 
Our general algorithm selects one of the identified bridges.
Each node in the bridge indicates one of the bridge connected components. 
Thus, we determine a new edge that connects both components which in turn do not include either of the nodes of the bridge.
Subsequently, we start over with the next bridge connecting two remaining bridge-connected components. 
The algorithm repeats the process until there are no more bridge-connected components left. 
A special case that needs to be treated before running the general algorithm is the appearance of {\itshape bridge paths}. 
We treat those first to prevent the general algorithm from infinitely looping. 
In such a case, we start at one end of the bridge path $P$ incorporating the nodes $v_s$, $v_{s+1}$, \ldots, $v_{s+k}$ and the edges $\{\{v_s, v_{s+1}\},\{v_{s+1}, v_{s+2}\},\ldots,\{v_{s+k-1},v_{s+k}\}\}$ of the graph $G=(V,E)$. 
Starting at $v_s$ of the bridge path, we add an edge to the graph from $v_s$ to the next but one node $v_{s+2}$. 
This procedure has to be repeated for each node except the nodes $v_{k-1}$ and $v_{k}$.
After applying this procedure, all bridge paths have been eliminated from the graph and the general algorithm to remove the bridges of the graph can be executed.

{\bfseries Average Node Degree:} The algorithm will remove edges from the graph until a desired average node degree has been achieved.
We have chosen the edge length as decisive property to select the edges to be removed, since in WSNs a connection between nodes that are further apart is less likely.
To accomplish this, the algorithm selects edges by different criteria.
The selection can be done randomly, with the probability of an edge being removed weighted by its length and a given exponent or in order by edge length.
As additional conditions, we can exclude edges that, if removed, would cause the graph to become disconnected or introduce bridges within it.

\section{Empirical Test Setup}\label{sec:empirical_setup}
To evaluate the computability of optimal and maximal $n$-soft domatic partitions for reasonably sized large-scale static homogeneous WSNs, we outline the details of our empirical test setup.
The corresponding graphs are created by the proposed $\lambda$-precision UDG generator using the seeds depicted in Subsection \ref{sec:gen_seeds} and the associated Table \ref{tbl:test}.
We have chosen graphs with a number of nodes $|V|$ starting from $20$ to $300$ in steps of $20$.
Only connected $\lambda$-precision UDGs created by our graph generator are accepted in our test setup.
In case a generated graph is not connected, we discard it and repeat the generation process for the given parameters until the desired number of connected $\lambda$-precision UDGs has been reached.
After successfully generating $20$ connected graphs for each row of parameter combinations in Table \ref{tbl:test}, we duplicate the complete set of graphs once for a second test setup.
The original set of graphs $SG_1$ is then adapted to approach the expected average node degree $\deg_\text{exp}$ by successive removal of edges.
The algorithm used to adapt the graphs and to reach the desired average node degree is described in Subsection \ref{sec:graph_adapt}.
To adjust the average node degree, we take the squared edge length of each edge that does not disconnect the graph as weight.
The squaring gives longer edges a higher priority to be selected in the process.
Then, edges are removed iteratively and by chance based on their respective given weight until the average node degree $\deg_\text{avg}$ reaches the desired expected average node degree $\deg_\text{exp}$.
The graphs in the duplicated set $SG_2$ are modified by removing all bridges as described in Subsection \ref{sec:graph_adapt}. 
Afterwards, we ensure that the $\deg_\text{exp}$ in the table row associated with the graph is reached as described for $SG_1$ but without the risk of creating new bridges. 
The set of graphs $SG_2$ is created to evaluate whether small topological properties like for a graph to be bridge-free in our given set of graphs directly affects the computability or quality of results of our partitioning schemes.
Finally, we compute for all graphs the optimal and maximal $n$-soft domatic partitions for $n\in \{3,4,5\}$.
For this purpose, the $0-1$ LPs have been implemented using Pyomo \cite{hart2011pyomo} and they are computed using the mathematical programming solver Gurobi \cite{gurobi}. 
In the last step, we evaluate the results via Python.
Therefore, we track the wallclock times given by Gurobi.
In addition, we count the number of missing coverages $e_\text{miss\_cov}$ introduced in Equation \eqref{eq:miss_cov} as well as the incompletely covered nodes $e_\text{inc\_nodes}$ expressed in Equation \eqref{eq:inc_nodes}. 
The time limit for Gurobi to solve a given LP on a given graph is set to $1200$ seconds on a system with two {\itshape Intel$^\text{\textregistered}$ Xeon$^\text{\textregistered}$ Gold 6248R} as central processing units and $256$ GB of random access memory.

\section{Results and Evaluation}\label{sec:res_eval}
Here, we evaluate the computation results of the optimal and maximal $n$-soft domatic partitions with $n\in \{3,4,5\}$ and for $2400$ different $\lambda$-precision UDGs divided into two test sets $SG_1$ and $SG_2$ as described in the previous section.

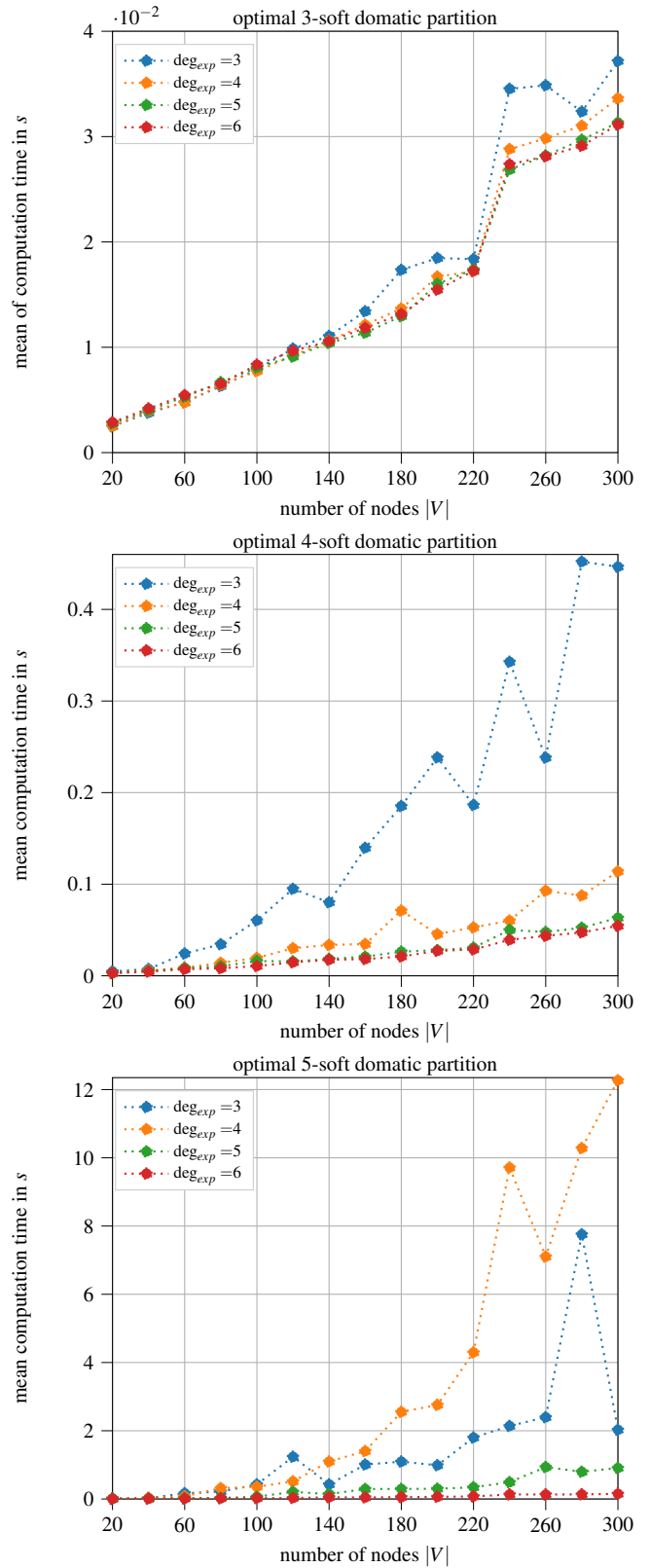
\begin{figure}
	\centering
\begin{tikzpicture}
\definecolor{crimson2143940}{RGB}{214,39,40}
\definecolor{darkgray176}{RGB}{176,176,176}
\definecolor{darkorange25512714}{RGB}{255,127,14}
\definecolor{forestgreen4416044}{RGB}{44,160,44}
\definecolor{gainsboro229}{RGB}{229,229,229}
\definecolor{lightgray204}{RGB}{204,204,204}
\definecolor{steelblue31119180}{RGB}{31,119,180}

\begin{axis}[
legend style={
	fill opacity=1, 
	draw opacity=1, 
	text opacity=1, 
	at={(0.005,0.97)},
	anchor=north west,
	draw=lightgray204,
	nodes={scale=0.8, transform shape}
},
tick align=outside,
tick pos=left,
title style={yshift=-8pt},
title={optimal 3-soft domatic partition},
x grid style={darkgray176},
xlabel={number of nodes \(\displaystyle |V|\)},
xmajorgrids,
xmin=20, xmax=300,
xtick style={color=black},
xtick={20, 60, 100, 140, 180, 220, 260, 300},
y grid style={darkgray176},
ylabel={mean of computation time in \(\displaystyle s\)},
ymajorgrids,
ymin=0, ymax=0.04,
ytick style={color=black},
style={/pgf/number format/fixed}
]

\addplot [thick, steelblue31119180, dotted, mark=*]
table {%
20 0.00249197483062744
40 0.00378146171569824
60 0.00477365255355835
80 0.00631090402603149
100 0.00789241790771484
120 0.00987063646316528
140 0.0110679745674133
160 0.0134113073348999
180 0.0173452496528625
200 0.018467903137207
220 0.018363618850708
240 0.0345288872718811
260 0.0348653674125671
280 0.0323670268058777
300 0.0371704578399658
};
\addlegendentry{$\deg_{exp}=$3}
\addplot [thick, darkorange25512714, dotted, mark=*]
table {%
20 0.0025012731552124
40 0.00392756462097168
60 0.00474729537963867
80 0.00642237663269043
100 0.00773278474807739
120 0.00920650959014893
140 0.0106282949447632
160 0.0121333718299866
180 0.0136736392974854
200 0.0167043328285217
220 0.0172057271003723
240 0.0288064241409302
260 0.029833984375
280 0.0310288548469543
300 0.0336383581161499
};
\addlegendentry{$\deg_{exp}=$4}
\addplot [thick, forestgreen4416044, dotted, mark=*]
table {%
20 0.00276203155517578
40 0.00401557683944702
60 0.00529978275299072
80 0.00668148994445801
100 0.0080793023109436
120 0.009100173649035
140 0.0104018568992615
160 0.0113720893859863
180 0.0129410028457642
200 0.0160166263580322
220 0.0173779487609863
240 0.0268582820892334
260 0.0281968355178833
280 0.0296975612640381
300 0.0313395142555237
};
\addlegendentry{$\deg_{exp}=$5}
\addplot [thick, crimson2143940, dotted, mark=*]
table {%
20 0.00288213491439819
40 0.00418953895568848
60 0.00545215606689453
80 0.00652980804443359
100 0.0083470344543457
120 0.00965758562088013
140 0.0105678081512451
160 0.0118348002433777
180 0.0131490349769592
200 0.0154636263847351
220 0.0172543883323669
240 0.0273655652999878
260 0.0281206607818604
280 0.0291190147399902
300 0.0311388969421387
};
\addlegendentry{$\deg_{exp}=$6}
\end{axis}

\end{tikzpicture}
\begin{tikzpicture}
\definecolor{crimson2143940}{RGB}{214,39,40}
\definecolor{darkgray176}{RGB}{176,176,176}
\definecolor{darkorange25512714}{RGB}{255,127,14}
\definecolor{forestgreen4416044}{RGB}{44,160,44}
\definecolor{gainsboro229}{RGB}{229,229,229}
\definecolor{lightgray204}{RGB}{204,204,204}
\definecolor{steelblue31119180}{RGB}{31,119,180}

\begin{axis}[
legend cell align={left},
legend style={
	fill opacity=1, 
	draw opacity=1, 
	text opacity=1, 
	at={(0.005,0.97)},
	anchor=north west,
	draw=lightgray204,
	nodes={scale=0.8, transform shape}
},
tick align=outside,
tick pos=left,
title style={yshift=-8pt},
title={optimal 4-soft domatic partition},
x grid style={darkgray176},
xlabel={number of nodes \(\displaystyle |V|\)},
xmajorgrids,
xmin=20, xmax=300,
xtick style={color=black},
xtick={20, 60, 100, 140, 180, 220, 260, 300},
y grid style={darkgray176},
ylabel={mean computation time in \(\displaystyle s\)},
ymajorgrids,
ymin=0.00, ymax=0.46,
ytick style={color=black},
style={/pgf/number format/fixed}
]
\addplot [thick, steelblue31119180, dotted, mark=*]
table {%
20 0.00472623109817505
40 0.00735466480255127
60 0.0243144273757935
80 0.0343522548675537
100 0.0604203343391418
120 0.0948732495307922
140 0.0801615357398987
160 0.139599704742432
180 0.185310590267181
200 0.238309228420258
220 0.186427748203278
240 0.342753410339355
260 0.238129651546478
280 0.45209493637085
300 0.446553659439087
};
\addlegendentry{$\deg_{exp}=$3}
\addplot [thick, darkorange25512714, dotted, mark=*]
table {%
20 0.00299046039581299
40 0.00583211183547974
60 0.00868215560913086
80 0.0141334176063538
100 0.0194544911384583
120 0.0300884962081909
140 0.0336900472640991
160 0.0346366286277771
180 0.0711453914642334
200 0.0454451560974121
220 0.0527921438217163
240 0.060259735584259
260 0.092872154712677
280 0.0875398993492126
300 0.113945531845093
};
\addlegendentry{$\deg_{exp}=$4}
\addplot [thick, forestgreen4416044, dotted, mark=*]
table {%
20 0.00304497480392456
40 0.00511550903320312
60 0.0083932638168335
80 0.0103625774383545
100 0.016110360622406
120 0.0156390309333801
140 0.0183600306510925
160 0.0208682417869568
180 0.026364278793335
200 0.0279329538345337
220 0.0307166814804077
240 0.0500395894050598
260 0.0477768301963806
280 0.0524048089981079
300 0.0633179426193237
};
\addlegendentry{$\deg_{exp}=$5}
\addplot [thick, crimson2143940, dotted, mark=*]
table {%
20 0.0029538631439209
40 0.00451580286026001
60 0.00713167190551758
80 0.00815703868865967
100 0.0106170553910105
120 0.014663827419281
140 0.0174774050712585
160 0.0181976199150085
180 0.0210554003715515
200 0.0271612286567688
220 0.0287036776542664
240 0.0392426013946533
260 0.0435279369354248
280 0.0473247647285461
300 0.0545235157012939
};
\addlegendentry{$\deg_{exp}=$6}
\end{axis}

\end{tikzpicture}
\begin{tikzpicture}

\definecolor{crimson2143940}{RGB}{214,39,40}
\definecolor{darkgray176}{RGB}{176,176,176}
\definecolor{darkorange25512714}{RGB}{255,127,14}
\definecolor{forestgreen4416044}{RGB}{44,160,44}
\definecolor{gainsboro229}{RGB}{229,229,229}
\definecolor{lightgray204}{RGB}{204,204,204}
\definecolor{steelblue31119180}{RGB}{31,119,180}

\begin{axis}[
legend cell align={left},
legend style={
	fill opacity=1, 
	draw opacity=1, 
	text opacity=1, 
	at={(0.005,0.97)},
	anchor=north west,
	draw=lightgray204,
	nodes={scale=0.8, transform shape}
},
tick align=outside,
tick pos=left,
title style={yshift=-8pt},
title={optimal 5-soft domatic partition},
x grid style={darkgray176},
xlabel={number of nodes \(\displaystyle |V|\)},
xmajorgrids,
xmin=20, xmax=300,
xtick style={color=black},
xtick={20, 60, 100, 140, 180, 220, 260, 300},
y grid style={darkgray176},
ylabel={mean computation time in \(\displaystyle s\)},
ymajorgrids,
ymin=0, ymax=12.35,
ytick style={color=black},
style={/pgf/number format/fixed}
]
\addplot [thick, steelblue31119180, dotted, mark=*]
table {%
20 0.0144946694374084
40 0.0240044355392456
60 0.167525577545166
80 0.215345096588135
100 0.425712871551514
120 1.23647688627243
140 0.426715576648712
160 1.00814572572708
180 1.0945925951004
200 0.987664449214935
220 1.80005118846893
240 2.13955053091049
260 2.39568566083908
280 7.75577971935272
300 2.0288699388504
};
\addlegendentry{$\deg_{exp}=$3}
\addplot [thick, darkorange25512714, dotted, mark=*]
table {%
20 0.00702849626541138
40 0.0286633849143982
60 0.0781733989715576
80 0.316292238235474
100 0.355808544158936
120 0.514993166923523
140 1.09683212041855
160 1.40294815301895
180 2.5515688419342
200 2.75634763240814
220 4.29662646055222
240 9.71328996419907
260 7.10247659683228
280 10.2836630940437
300 12.2717106819153
};
\addlegendentry{$\deg_{exp}=$4}
\addplot [thick, forestgreen4416044, dotted, mark=*]
table {%
20 0.00600054264068603
40 0.0102184295654297
60 0.0225603699684143
80 0.0298203110694885
100 0.0662891507148743
120 0.208377242088318
140 0.143903565406799
160 0.291485333442688
180 0.293760168552399
200 0.29921441078186
220 0.342956161499023
240 0.491958487033844
260 0.935448455810547
280 0.796795916557312
300 0.908484160900116
};
\addlegendentry{$\deg_{exp}=$5}
\addplot [thick, crimson2143940, dotted, mark=*]
table {%
20 0.00404039621353149
40 0.008652663230896
60 0.0118186116218567
80 0.0188933491706848
100 0.0237881660461426
120 0.0282766222953796
140 0.0502007365226746
160 0.0442652702331543
180 0.0536921143531799
200 0.0625409126281738
220 0.0748098015785217
240 0.133429539203644
260 0.130683350563049
280 0.134118068218231
300 0.150491857528687
};
\addlegendentry{$\deg_{exp}=$6}
\end{axis}

\end{tikzpicture}
	\caption{Test results of the mean of the computation time in seconds $s$ subject to the number of nodes $|V|$ of given $\lambda$-precision UDGs necessary to determine optimal $n$-soft domatic partitions within a time limit of $1200$ $s$.}
	\label{fig:res_opt}
\end{figure}

First, we start with solely discussing the results computed on $SG_1$. 
In Fig. \ref{fig:res_opt}, we evaluate the mean of the computation time of the optimal $3$, $4$ and $5$-soft domatic partitions in dependence on the number of nodes $|V|$ of the given graphs.
The colours of the respective curves represent the expected average node degree of the given graphs. 
The dotted lines in between the drawn data points are added exclusively to improve the readability of the plots.


All plots in Fig. \ref{fig:res_opt} show a similar behaviour.
The computation time seems to increase for graphs with an average node degree lower or equal to the partition size of a graph.
The increase is potentially caused by the implicit increase of a number of missing coverages $\overline{e_\text{miss\_cov}}$ in those graphs as illustrated in Tables \ref{tbl:opt_max_3}, \ref{tbl:opt_max_4}, \ref{tbl:opt_max_5} and \ref{tbl:opt_max_6}.
For all graphs the $n$-soft domatic partitions could be solved to optimality.
The required computation time remained below $13$ seconds for each graph and determined partition.
Therefore, even larger graphs should be solvable in reasonable time.
Deviations and jumps between node numbers can potentially be attributed to the limited number of test cases, which influence the test result through individual outliers.
Such jumps can be observed, for example, in the figure for the optimal $5$-soft domatic partitioning in Figure \ref{fig:res_opt}.
There, the curve for graphs with average node degree $5$ between $260$ and $300$ nodes includes an unexpected increase in computation time for graphs with $280$ nodes.

Overall, the number of test cases seems to be sufficient to derive a trend towards the required computation time.
Moreover, the results show that the computability of the optimal $n$-soft domatic partitions for $n\in \{3,4,5\}$ is given for the set of graphs $SG_1$.

\begin{figure}
	\centering
\begin{tikzpicture}
\definecolor{crimson2143940}{RGB}{214,39,40}
\definecolor{darkgray176}{RGB}{176,176,176}
\definecolor{darkorange25512714}{RGB}{255,127,14}
\definecolor{forestgreen4416044}{RGB}{44,160,44}
\definecolor{gainsboro229}{RGB}{229,229,229}
\definecolor{lightgray204}{RGB}{204,204,204}
\definecolor{steelblue31119180}{RGB}{31,119,180}

\begin{axis}[
legend cell align={left},
legend style={
	fill opacity=1, 
	draw opacity=1, 
	text opacity=1, 
	at={(0.005,0.97)},
	anchor=north west,
	draw=lightgray204,
	nodes={scale=0.8, transform shape}
},
tick align=outside,
tick pos=left,
title style={yshift=-8pt},
title={maximal 3-soft domatic partition},
x grid style={darkgray176},
xlabel={number of nodes \(\displaystyle |V|\)},
xmajorgrids,
xmin=20, xmax=300,
xtick style={color=black},
xtick={20, 60, 100, 140, 180, 220, 260, 300},
y grid style={darkgray176},
ylabel={mean computation time in \(\displaystyle s\)},
ymajorgrids,
ymin=0, ymax=0.33,
ytick style={color=black}
]
\addplot [thick, steelblue31119180, dotted, mark=*]
table {%
20 0.0126965641975403
40 0.0250478744506836
60 0.0361930847167969
80 0.0545272707939148
100 0.0630501389503479
120 0.0803331255912781
140 0.112751257419586
160 0.127487945556641
180 0.146381139755249
200 0.203624141216278
220 0.193129467964172
240 0.248497879505157
260 0.252159547805786
280 0.251059353351593
300 0.328009414672852
};
\addlegendentry{$\deg_{exp}=$3}
\addplot [thick, darkorange25512714, dotted, mark=*]
table {%
20 0.0120586395263672
40 0.0190291285514832
60 0.0234086394309998
80 0.0301391959190369
100 0.033294939994812
120 0.0407209157943726
140 0.0455513715744019
160 0.0546292185783386
180 0.0607227563858032
200 0.0719281077384949
220 0.0738641500473022
240 0.0951795101165771
260 0.107683491706848
280 0.112514734268188
300 0.13326427936554
};
\addlegendentry{$\deg_{exp}=$4}
\addplot [thick, forestgreen4416044, dotted, mark=*]
table {%
20 0.00980569124221802
40 0.0167673230171204
60 0.0209755778312683
80 0.0239652872085571
100 0.028077495098114
120 0.0321688175201416
140 0.0367974400520325
160 0.0402623891830444
180 0.0452333688735962
200 0.049996542930603
220 0.0548749685287476
240 0.0693184614181519
260 0.0806342005729675
280 0.0841676115989685
300 0.0903956294059753
};
\addlegendentry{$\deg_{exp}=$5}
\addplot [thick, crimson2143940, dotted, mark=*]
table {%
20 0.00977218778509843
40 0.0153831958770752
60 0.0204619646072388
80 0.0227081179618835
100 0.0256590604782104
120 0.0296229720115662
140 0.0333515286445618
160 0.0368239283561707
180 0.0433685898780823
200 0.0482110500335693
220 0.0578404188156128
240 0.0705667972564697
260 0.0772366046905518
280 0.0773211240768433
300 0.0796231269836426
};
\addlegendentry{$\deg_{exp}=$6}
\end{axis}

\end{tikzpicture}
\begin{tikzpicture}
\definecolor{crimson2143940}{RGB}{214,39,40}
\definecolor{darkgray176}{RGB}{176,176,176}
\definecolor{darkorange25512714}{RGB}{255,127,14}
\definecolor{forestgreen4416044}{RGB}{44,160,44}
\definecolor{gainsboro229}{RGB}{229,229,229}
\definecolor{lightgray204}{RGB}{204,204,204}
\definecolor{steelblue31119180}{RGB}{31,119,180}

\begin{axis}[
legend cell align={left},
legend style={
	fill opacity=1, 
	draw opacity=1, 
	text opacity=1, 
	at={(0.005,0.97)},
	anchor=north west,
	draw=lightgray204,
	nodes={scale=0.8, transform shape}
},
tick align=outside,
tick pos=left,
title style={yshift=-8pt},
title={maximal 4-soft domatic partition},
x grid style={darkgray176},
xlabel={number of nodes \(\displaystyle |V|\)},
xmajorgrids,
xmin=20, xmax=300,
xtick style={color=black},
xtick={20, 60, 100, 140, 180, 220, 260, 300},
y grid style={darkgray176},
ylabel={mean computation time in \(\displaystyle s\)},
ymajorgrids,
ymin=0.00, ymax=2.1,
ytick style={color=black}
]
\addplot [thick, steelblue31119180, dotted, mark=*]
table {%
20 0.0336341619491577
40 0.110609781742096
60 0.22781697511673
80 0.374657130241394
100 0.47957456111908
120 0.618183207511902
140 0.697916984558105
160 0.969267606735229
180 1.14025593996048
200 1.23200290203094
220 1.47879753112793
240 1.57294626235962
260 1.65336742401123
280 2.06230248212814
300 1.98688941001892
};
\addlegendentry{$\deg_{exp}=$3}
\addplot [thick, darkorange25512714, dotted, mark=*]
table {%
20 0.0215333223342896
40 0.0532344937324524
60 0.120413768291473
80 0.190379309654236
100 0.255233061313629
120 0.353206288814545
140 0.413982677459717
160 0.476885855197906
180 0.711905574798584
200 0.680944907665253
220 0.746794748306274
240 0.862359130382538
260 1.02181921005249
280 1.09015208482742
300 1.37753649950027
};
\addlegendentry{$\deg_{exp}=$4}
\addplot [thick, forestgreen4416044, dotted, mark=*]
table {%
20 0.0165978908538818
40 0.0330304622650146
60 0.0485225439071655
80 0.0679478764533997
100 0.105861783027649
120 0.127816164493561
140 0.133438574640374
160 0.182111704349518
180 0.253075635433197
200 0.251438105106354
220 0.283178949356079
240 0.359677588939667
260 0.363655161857605
280 0.405377531051636
300 0.469977355003357
};
\addlegendentry{$\deg_{exp}=$5}
\addplot [thick, crimson2143940, dotted, mark=*]
table {%
20 0.0141908049583435
40 0.0254233717918396
60 0.0295594692230225
80 0.0396919727325439
100 0.0489062786102295
120 0.0625555038452148
140 0.0736903667449951
160 0.0798800349235535
180 0.107440459728241
200 0.114591300487518
220 0.135694169998169
240 0.172085309028625
260 0.164751958847046
280 0.226295757293701
300 0.200979566574097
};
\addlegendentry{$\deg_{exp}=$6}
\end{axis}

\end{tikzpicture}
\begin{tikzpicture}
\definecolor{crimson2143940}{RGB}{214,39,40}
\definecolor{darkgray176}{RGB}{176,176,176}
\definecolor{darkorange25512714}{RGB}{255,127,14}
\definecolor{forestgreen4416044}{RGB}{44,160,44}
\definecolor{gainsboro229}{RGB}{229,229,229}
\definecolor{lightgray204}{RGB}{204,204,204}
\definecolor{steelblue31119180}{RGB}{31,119,180}

\begin{axis}[
legend cell align={left},
legend style={
	fill opacity=1, 
	draw opacity=1, 
	text opacity=1, 
	at={(0.005,0.97)},
	anchor=north west,
	draw=lightgray204,
	nodes={scale=0.8, transform shape}
},
tick align=outside,
tick pos=left,
title style={yshift=-8pt},
title={maximal 5-soft domatic partition},
x grid style={darkgray176},
xlabel={number of nodes \(\displaystyle |V|\)},
xmajorgrids,
xmin=20, xmax=300,
xtick style={color=black},
xtick={20, 60, 100, 140, 180, 220, 260, 300},
y grid style={darkgray176},
ylabel={mean computation time in \(\displaystyle s\)},
ymajorgrids,
ymin=0.00, ymax=11.00,
ytick style={color=black}
]
\addplot [thick, steelblue31119180, dotted, mark=*]
table {%
20 0.0554861068725586
40 0.151774871349335
60 0.266283917427063
80 0.386484313011169
100 0.587322056293488
120 0.76153074502945
140 0.835177576541901
160 1.10459694862366
180 1.20101919174194
200 1.29929566383362
220 1.59152394533157
240 1.76630121469498
260 1.93932785987854
280 4.85298644304275
300 1.88427324295044
};
\addlegendentry{$\deg_{exp}=$3}
\addplot [thick, darkorange25512714, dotted, mark=*]
table {%
20 0.047821581363678
40 0.176203095912933
60 0.392671656608582
80 0.895702350139618
100 1.44232459068298
120 1.6281699180603
140 3.18051242828369
160 3.46698622703552
180 4.05460915565491
200 4.76363067626953
220 6.29271011352539
240 10.0264775633812
260 10.5273764252663
280 10.6284862756729
300 10.9362516760826
};
\addlegendentry{$\deg_{exp}=$4}
\addplot [thick, forestgreen4416044, dotted, mark=*]
table {%
20 0.0319199323654175
40 0.117622029781342
60 0.27479236125946
80 0.454770374298096
100 0.72408903837204
120 1.01474485397339
140 1.34815535545349
160 2.07440793514252
180 1.9356564283371
200 2.43222681283951
220 3.17006078958511
240 3.63761446475983
260 4.74959651231766
280 5.23190792798996
300 6.02413995265961
};
\addlegendentry{$\deg_{exp}=$5}
\addplot [thick, crimson2143940, dotted, mark=*]
table {%
20 0.0229902386665344
40 0.064396858215332
60 0.122497248649597
80 0.184614455699921
100 0.26883362531662
120 0.394019091129303
140 0.54211882352829
160 0.579523229598999
180 0.683251357078552
200 0.794084668159485
220 0.807068991661072
240 1.08742390871048
260 1.18843861818314
280 1.28567267656326
300 1.44353712797165
};
\addlegendentry{$\deg_{exp}=$6}
\end{axis}

\end{tikzpicture}
	\caption{Test results of the mean of the computation time in seconds $s$ necessary to determine maximal $n$-soft domatic partitions in dependence of the number of nodes $|V|$ of $\lambda$-precision UDGs within a time limit of $1200$ $s$.}
	\label{fig:res_max}
\end{figure}
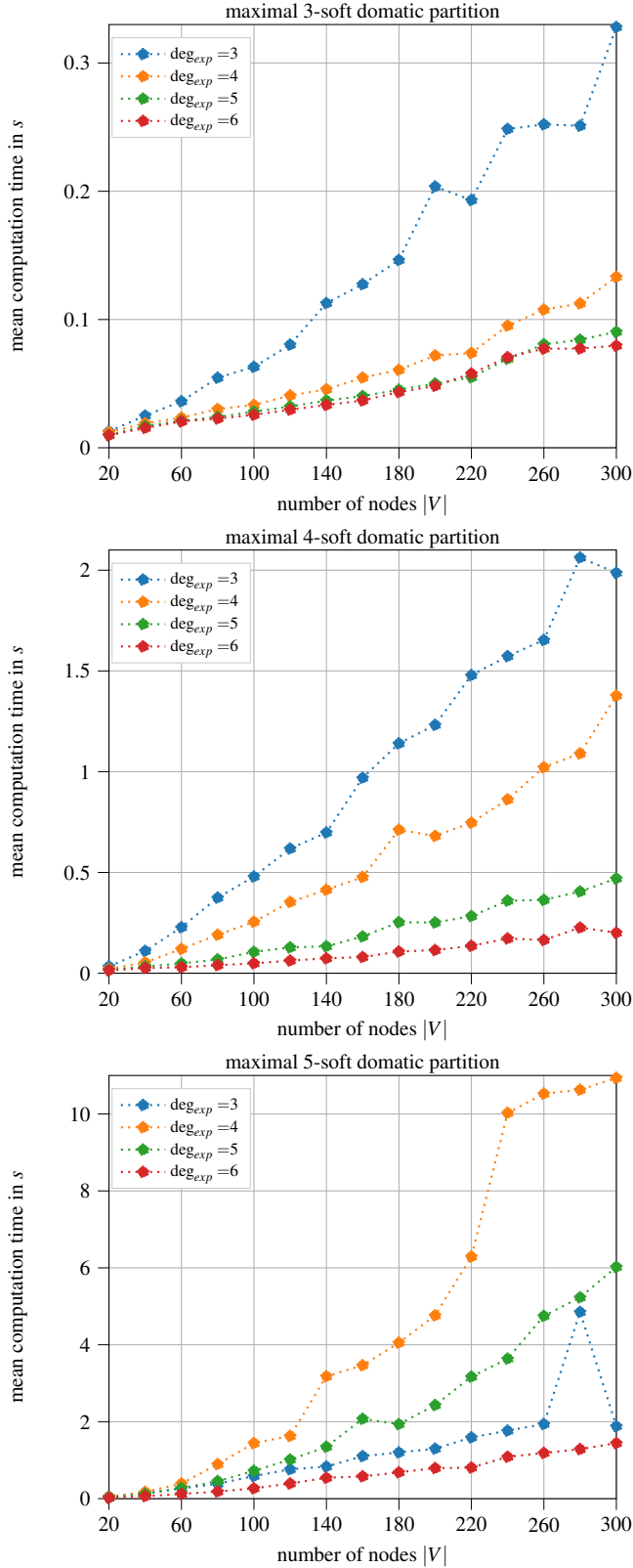

Fig. \ref{fig:res_max} shows the results for the computation of the maximal $3$, $4$ and $5$-soft domatic partitions of $SG_1$.
The diagrams are structured in the same way as the diagrams from Fig. \ref{fig:res_opt}.
The results in this figure are also subject to empirical fluctuations, particularly due to the limited number of test cases.

The topmost diagram of Fig. \ref{fig:res_max} displays the results of the maximal $3$-soft domatic partition. 
It shows that the mean of the computation time increases with an increasing number of nodes in the graph.
Furthermore, we can see that the increase of the average node degree comes with a decrease of the mean of the computation time. 
This can be a consequence of the decreasing number of nodes that do not cause any coverage errors and therefore contribute to the optimality result of the LPs.
A similar effect becomes visible in the results of the computation of the maximal $4$-soft domatic partition in Fig. \ref{fig:res_max} as well.
The observed patterns in computation time for maximal $3$ and $4$-soft domatic partitions align with the trends observed in the optimal $3$ and $4$-soft domatic partitions, as depicted in Fig. \ref{fig:res_opt}. 
These trends indicate an increase in computation time for graphs where the average node degrees are equal to or lower than the computed partition size.
The trend continues in general for the computation time of the optimal and maximal $5$-soft domatic partitions while the order of plots for graphs with average node degrees greater and equal to the partition changes compared to the other graphs.
In a broader sense, the consistent trend persists regarding the computation time for both the optimal and maximal $5$-soft domatic partitions. 
However, there is a variation in the arrangement of plots concerning graphs with average node degrees equal to or greater than the partition size, distinct from the other graph scenarios.
Overall, an observable pattern emerges where an increasing average node degree relative to the partition size notably rises the mean computation time.
All maximal $3$, $4$ and $5$-soft domatic partitions have been computed optimally within the given time limit of $1200$ seconds.

\begin{figure}
\begin{tikzpicture}

\definecolor{crimson2143940}{RGB}{214,39,40}
\definecolor{darkgray176}{RGB}{176,176,176}
\definecolor{darkorange25512714}{RGB}{255,127,14}
\definecolor{forestgreen4416044}{RGB}{44,160,44}
\definecolor{gainsboro229}{RGB}{229,229,229}
\definecolor{lightgray204}{RGB}{204,204,204}
\definecolor{steelblue31119180}{RGB}{31,119,180}

\begin{axis}[
legend style={
	fill opacity=1, 
	draw opacity=1, 
	text opacity=1, 
	at={(0.005,0.97)},
	anchor=north west,
	draw=lightgray204,
	nodes={scale=0.8, transform shape}
},
xtick style={color=black},
tick align=outside,
tick pos=left,
title style={yshift=-8pt},
title={optimal and maximal 3-soft domatic partition},
x grid style={darkgray176},
xlabel={number of nodes \(\displaystyle |V|\)},
xmajorgrids,
xmin=20, xmax=300,
xtick style={color=black},
xtick={20, 60, 100, 140, 180, 220, 260, 300},
y grid style={darkgray176},
ylabel={mean number of incompletely covered nodes},
ymajorgrids,
ymin=0, ymax=24,
ytick style={color=black},
style={/pgf/number format/fixed}
]
\addplot [thick, steelblue31119180, dashed, mark=*]
table {%
20 1.4
40 3.65
60 5.25
80 6.55
100 7
120 7.9
140 11.1
160 12.15
180 11.9
200 15
220 16.3
240 16.05
260 17.9
280 17.7
300 23.25
};
\addlegendentry{$\deg_{exp}=$3, opt}
\addplot [thick, steelblue31119180, dotted, mark=*]
table {%
20 1.4
40 3.65
60 5.25
80 6.55
100 7
120 7.85
140 11.1
160 12.15
180 11.9
200 15
220 16.3
240 16.05
260 17.9
280 17.7
300 23.25
};
\addlegendentry{$\deg_{exp}=$3, max}
\addplot [thick, darkorange25512714, dashed, mark=*]
table {%
20 0.75
40 1.2
60 1.5
80 2.35
100 2.25
120 3.25
140 3.2
160 3.75
180 3.7
200 4.45
220 3.8
240 5.1
260 4.7
280 4.95
300 6.5
};
\addlegendentry{$\deg_{exp}=$4, opt}
\addplot [thick, darkorange25512714, dotted, mark=*]
table {%
20 0.75
40 1.2
60 1.5
80 2.35
100 2.25
120 3.25
140 3.2
160 3.75
180 3.7
200 4.45
220 3.8
240 5.1
260 4.7
280 4.95
300 6.5
};
\addlegendentry{$\deg_{exp}=$4, max}
\addplot [thick, forestgreen4416044, dashed, mark=*]
table {%
20 0.2
40 0.5
60 0.6
80 0.75
100 1.05
120 1.10526315789474
140 1.35
160 0.95
180 1.4
200 1.25
220 1.4
240 1.6
260 1.75
280 1.75
300 2
};
\addlegendentry{$\deg_{exp}=$5, opt}
\addplot [thick, forestgreen4416044, dotted, mark=*]
table {%
20 0.2
40 0.5
60 0.6
80 0.75
100 1.05
120 1.05
140 1.35
160 0.95
180 1.4
200 1.25
220 1.4
240 1.6
260 1.75
280 1.75
300 2
};
\addlegendentry{$\deg_{exp}=$5, max}
\addplot [thick, crimson2143940, dashed, mark=*]
table {%
20 0.1
40 0.15
60 0.3
80 0.3
100 0.2
120 0.35
140 0.5
160 0.45
180 0.4
200 0.8
220 0.85
240 0.8
260 0.9
280 0.9
300 0.8
};
\addlegendentry{$\deg_{exp}=$6, opt}
\addplot [thick, crimson2143940, dotted, mark=*]
table {%
20 0.105263157894737
40 0.15
60 0.3
80 0.3
100 0.2
120 0.35
140 0.5
160 0.45
180 0.4
200 0.8
220 0.85
240 0.8
260 0.9
280 0.9
300 0.8
};
\addlegendentry{$\deg_{exp}=$6, max}
\end{axis}

\end{tikzpicture}
\begin{tikzpicture}

\definecolor{crimson2143940}{RGB}{214,39,40}
\definecolor{darkgray176}{RGB}{176,176,176}
\definecolor{darkorange25512714}{RGB}{255,127,14}
\definecolor{forestgreen4416044}{RGB}{44,160,44}
\definecolor{gainsboro229}{RGB}{229,229,229}
\definecolor{lightgray204}{RGB}{204,204,204}
\definecolor{steelblue31119180}{RGB}{31,119,180}

\begin{axis}[
legend style={
	fill opacity=1, 
	draw opacity=1, 
	text opacity=1, 
	at={(0.005,0.97)},
	anchor=north west,
	draw=lightgray204,
	nodes={scale=0.8, transform shape}
},
xtick style={color=black},
tick align=outside,
tick pos=left,
title style={yshift=-8pt},
title={optimal and maximal 4-soft domatic partition},
x grid style={darkgray176},
xlabel={number of nodes \(\displaystyle |V|\)},
xmajorgrids,
xmin=20, xmax=300,
xtick style={color=black},
xtick={20, 60, 100, 140, 180, 220, 260, 300},
y grid style={darkgray176},
ylabel={mean number of incompletely covered nodes},
ymajorgrids,
ymin=0, ymax=98,
ytick style={color=black},
style={/pgf/number format/fixed}
]
\addplot [thick, steelblue31119180, dashed, mark=*]
table {%
20 6.6
40 11.8
60 18.95
80 24.65
100 33.3
120 35.5
140 43.1
160 49.6
180 53.35
200 61.05
220 61.15
240 69.25
260 74.55
280 76.2
300 94.7
};
\addlegendentry{$\deg_{exp}=$3, opt}
\addplot [thick, steelblue31119180, dotted, mark=*]
table {%
20 6.6
40 11.8
60 18.85
80 24.6
100 33.2
120 35.3
140 43
160 49.6
180 53.1
200 60.95
220 61.1
240 68.95
260 74.4
280 76.1
300 94.5
};
\addlegendentry{$\deg_{exp}=$3, max}
\addplot [thick, darkorange25512714, dashed, mark=*]
table {%
20 3.25
40 6.55
60 7.55
80 11
100 13
120 14.9
140 16.3
160 21
180 19.7
200 22.4
220 23
240 28.6
260 28.4
280 29.8
300 35.4
};
\addlegendentry{$\deg_{exp}=$4, opt}
\addplot [thick, darkorange25512714, dotted, mark=*]
table {%
20 3.25
40 6.55
60 7.55
80 10.95
100 12.95
120 14.8
140 16.3
160 21
180 19.65
200 22.35
220 23
240 28.6
260 28.4
280 29.7
300 35.35
};
\addlegendentry{$\deg_{exp}=$4, max}
\addplot [thick, forestgreen4416044, dashed, mark=*]
table {%
20 1.55
40 2.85
60 3.65
80 5.4
100 5.75
120 5.7
140 6.8
160 7.6
180 8.9
200 9
220 9.25
240 10.55
260 11.65
280 13.1
300 13.35
};
\addlegendentry{$\deg_{exp}=$5, opt}
\addplot [thick, forestgreen4416044, dotted, mark=*]
table {%
20 1.55
40 2.85
60 3.65
80 5.4
100 5.75
120 5.7
140 6.57894736842105
160 7.6
180 8.9
200 9
220 9.25
240 10.55
260 11.65
280 13.1
300 13.35
};
\addlegendentry{$\deg_{exp}=$5, max}
\addplot [thick, crimson2143940, dashed, mark=*]
table {%
20 0.65
40 1.4
60 1.85
80 2.05
100 2.5
120 3.15
140 3.15
160 3.25
180 3.7
200 3.75
220 4.6
240 4.65
260 5.25
280 5.3
300 6.05
};
\addlegendentry{$\deg_{exp}=$6, opt}
\addplot [thick, crimson2143940, dotted, mark=*]
table {%
20 0.65
40 1.4
60 1.85
80 2.05
100 2.5
120 3.15
140 3.15
160 3.25
180 3.7
200 3.75
220 4.6
240 4.65
260 5.25
280 5.3
300 6.05
};
\addlegendentry{$\deg_{exp}=$6, max}
\end{axis}

\end{tikzpicture}
\begin{tikzpicture}

\definecolor{crimson2143940}{RGB}{214,39,40}
\definecolor{darkgray176}{RGB}{176,176,176}
\definecolor{darkorange25512714}{RGB}{255,127,14}
\definecolor{forestgreen4416044}{RGB}{44,160,44}
\definecolor{gainsboro229}{RGB}{229,229,229}
\definecolor{lightgray204}{RGB}{204,204,204}
\definecolor{steelblue31119180}{RGB}{31,119,180}

\begin{axis}[
legend style={
	fill opacity=1, 
	draw opacity=1, 
	text opacity=1, 
	at={(0.005,0.97)},
	anchor=north west,
	draw=lightgray204,
	nodes={scale=0.8, transform shape}
},
xtick style={color=black},
tick align=outside,
tick pos=left,
title style={yshift=-8pt},
title={optimal and maximal 5-soft domatic partition},
x grid style={darkgray176},
xlabel={number of nodes \(\displaystyle |V|\)},
xmajorgrids,
xmin=20, xmax=300,
xtick style={color=black},
xtick={20, 60, 100, 140, 180, 220, 260, 300},
y grid style={darkgray176},
ylabel={mean number of incompletely covered nodes},
ymajorgrids,
ymin=-8.28, ymax=295.98,
ytick style={color=black},
style={/pgf/number format/fixed}
]
\addplot [thick, steelblue31119180, dashed, mark=*]
table {%
20 12.7
40 25.5
60 37.05
80 49.9
100 64.3
120 77.7
140 90.05
160 104.6
180 114.35
200 125.75
220 135.25
240 151.5
260 160.3
280 170.85
300 194.6
};
\addlegendentry{$\deg_{exp}=$3, opt}
\addplot [thick, steelblue31119180, dotted, mark=*]
table {%
20 12.55
40 25.5
60 36.9
80 49.8
100 64.15
120 77.25
140 89.95
160 104.1
180 113.95
200 125.4
220 134.75
240 150.9
260 159.6
280 170.2
300 194.1
};
\addlegendentry{$\deg_{exp}=$3, max}
\addplot [thick, darkorange25512714, dashed, mark=*]
table {%
20 7.65
40 14.05
60 21.15
80 27.5
100 32.5
120 40.15
140 44.15
160 52.3
180 56.95
200 62.05
220 66.5
240 77.95
260 81.65
280 85.95
300 100.35
};
\addlegendentry{$\deg_{exp}=$4, opt}
\addplot [thick, darkorange25512714, dotted, mark=*]
table {%
20 7.65
40 14.05
60 21.05
80 27.3
100 32.45
120 39.95
140 44.05
160 52.15
180 56.6
200 61.85
220 66.25
240 77.75
260 81.25
280 86.85
300 99.9
};
\addlegendentry{$\deg_{exp}=$4, max}
\addplot [thick, forestgreen4416044, dashed, mark=*]
table {%
20 4.05
40 7.8
60 11.15
80 13.65
100 16.85
120 19.45
140 20.9
160 22.75
180 28.7
200 30.2
220 32.4
240 35.65
260 38.6
280 39.6
300 43.8
};
\addlegendentry{$\deg_{exp}=$5, opt}
\addplot [thick, forestgreen4416044, dotted, mark=*]
table {%
20 4.05
40 7.8
60 11.15
80 13.65
100 16.85
120 19.45
140 20.85
160 22.7
180 28.7
200 30.2
220 32.3
240 35.6
260 38.6
280 39.6
300 43.8
};
\addlegendentry{$\deg_{exp}=$5, max}
\addplot [thick, crimson2143940, dashed, mark=*]
table {%
20 2
40 4.9
60 6.5
80 7.5
100 8.8
120 10.8
140 11.55
160 11.2
180 14.15
200 14.2
220 16
240 18.45
260 17.1
280 19.95
300 21.55
};
\addlegendentry{$\deg_{exp}=$6, opt}
\addplot [thick, crimson2143940, dotted, mark=*]
table {%
20 2
40 4.9
60 6.5
80 7.5
100 8.8
120 10.8
140 11.55
160 11.2
180 14.15
200 14.2
220 16
240 18.45
260 17.1
280 19.95
300 21.55
};
\addlegendentry{$\deg_{exp}=$6, max}
\end{axis}

\end{tikzpicture}
	\caption{Arithmetic mean of the number of incompletely covered nodes in optimal and maximal $n$-soft domatic partitions subject to the number of nodes $|V|$ of the given $\lambda$-precision UDGs of optimal $n$-soft domatic partitions within a time limit of $1200$ $s$.}
	\label{fig:inc_cov_max_opt}
\end{figure}
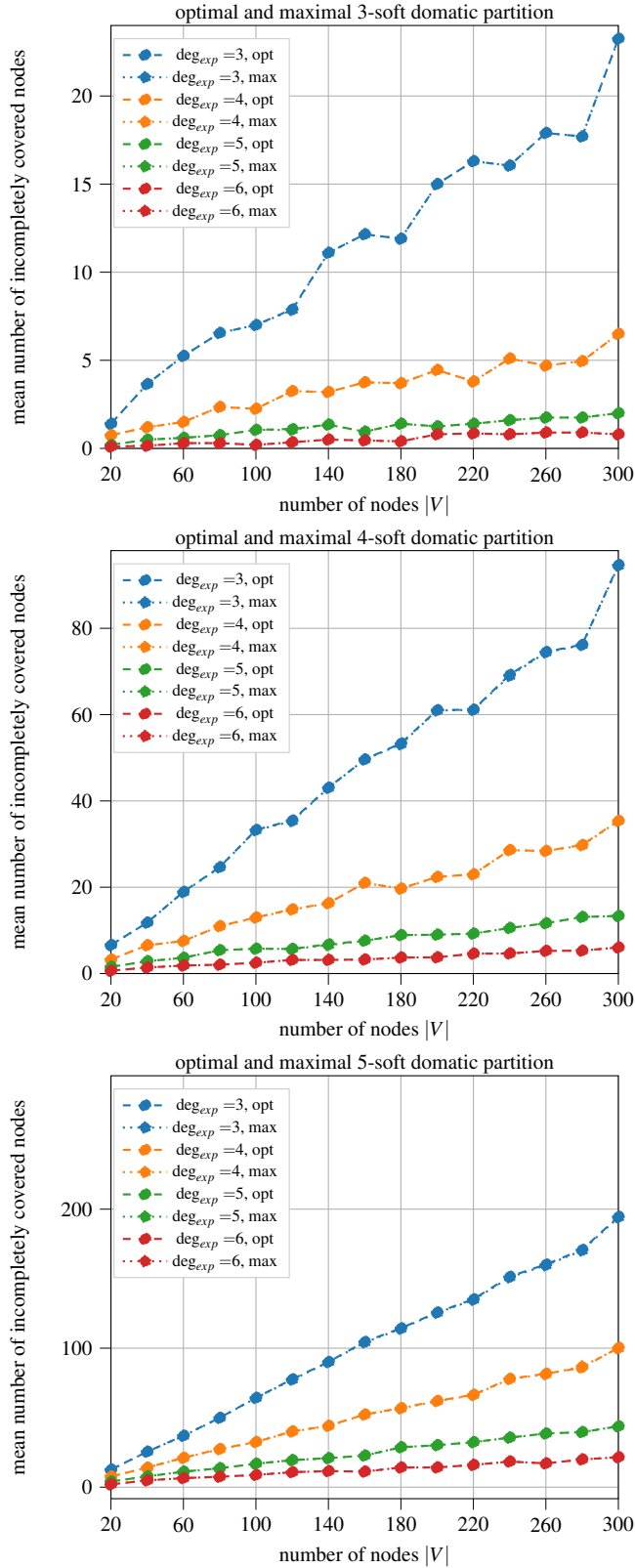

In Tables \ref{tbl:opt_max_3}, \ref{tbl:opt_max_4}, \ref{tbl:opt_max_5} and \ref{tbl:opt_max_6}, we compare and evaluate the cases in which for the given test setup and the set of graphs $SG_1$ and for the given set of parameters at least one solution has been computed optimally and one non-optimally within the given time limit for either optimal or maximal $n$-soft domatic partitions. 
\begin{table}[bh!]
	\centering
	\caption{Results for average node degree $\deg = 3$, partition sizes $n \in \{3, 4, 5\}$ and node numbers $|V|=\{20,40,\ldots, 300\}$ for optimal and maximal $n$-soft domatic partitions. Showing the mean of the number of missing coverages $\overline{e_\text{miss\_cov}}$ and number of incompletely covered nodes $\overline{e_\text{inc\_nodes}}$.}
	\label{tbl:opt_max_3}
	\scriptsize
	\rowcolors{2}{gray!25}{white}
	\begin{tabular}{rr|rr|rr}
		\specialrule{.5pt}{0pt}{.5pt}
		\raggedleft $n$ &
		\raggedleft $|V|$ &
		\raggedleft $\overline{e_\text{miss\_cov}}$ &
		\raggedleft $\overline{e_\text{inc\_nodes}}$ &
		\raggedleft $\overline{e_\text{miss\_cov}}$ &
		\raggedleft $\overline{e_\text{inc\_nodes}}$ \tabularnewline
		\specialrule{.5pt}{.5pt}{.5pt}
		  &     & \multicolumn{2}{c|}{opt $n$-soft domatic partition} & \multicolumn{2}{c}{max $n$-soft domatic partition}\\
		3 &  20 &   1.40 &   1.40 &   4.20 &   1.40 \\
		  &  40 &   3.65 &   3.65 &  10.95 &   3.65 \\
		  &  60 &   5.25 &   5.25 &  15.75 &   5.25 \\
		  &  80 &   6.55 &   6.55 &  19.65 &   6.55 \\
		  & 100 &   7.00 &   7.00 &  21.00 &   7.00 \\
		  & 120 &   7.90 &   7.90 &  23.55 &   7.85 \\
		  & 140 &  11.10 &  11.10 &  33.30 &  11.10 \\
		  & 160 &  12.15 &  12.15 &  36.45 &  12.15 \\
		  & 180 &  11.90 &  11.90 &  35.70 &  11.90 \\
		  & 200 &  15.00 &  15.00 &  45.00 &  15.00 \\
		  & 220 &  16.30 &  16.30 &  48.90 &  16.30 \\
		  & 240 &  16.05 &  16.05 &  48.15 &  16.05 \\
		  & 260 &  17.90 &  17.90 &  53.70 &  17.90 \\
		  & 280 &  17.70 &  17.70 &  53.10 &  17.70 \\
		  & 300 &  23.25 &  23.25 &  69.75 &  23.25 \\
		4 &  20 &   8.00 &   6.60 &  26.40 &   6.60 \\
		  &  40 &  15.45 &  11.80 &  47.20 &  11.80 \\
		  &  60 &  24.20 &  18.95 &  75.40 &  18.85 \\
		  &  80 &  31.15 &  24.65 &  98.40 &  24.60 \\
		  & 100 &  40.30 &  33.30 & 132.80 &  33.20 \\
		  & 120 &  43.35 &  35.50 & 141.20 &  35.30 \\
		  & 140 &  54.20 &  43.10 & 172.00 &  43.00 \\
		  & 160 &  61.70 &  49.60 & 198.40 &  49.60 \\
		  & 180 &  65.35 &  53.35 & 212.40 &  53.10 \\
		  & 200 &  76.05 &  61.05 & 243.80 &  60.95 \\
		  & 220 &  77.45 &  61.15 & 244.40 &  61.10 \\
		  & 240 &  85.40 &  69.25 & 275.80 &  68.95 \\
		  & 260 &  92.50 &  74.55 & 297.60 &  74.40 \\
		  & 280 &  93.90 &  76.20 & 304.40 &  76.10 \\
		  & 300 & 117.85 &  94.70 & 378.00 &  94.50 \\
		5 &  20 &  20.65 &  12.70 &  62.75 &  12.55 \\
		  &  40 &  40.95 &  25.50 & 127.50 &  25.50 \\
		  &  60 &  61.15 &  37.05 & 184.50 &  36.90 \\
		  &  80 &  81.10 &  49.90 & 249.00 &  49.80 \\
		  & 100 & 104.40 &  64.30 & 320.75 &  64.15 \\
		  & 120 & 120.50 &  77.70 & 386.25 &  77.25 \\
		  & 140 & 143.85 &  90.05 & 449.75 &  89.95 \\
		  & 160 & 165.80 & 104.60 & 520.50 & 104.10 \\
		  & 180 & 178.95 & 114.35 & 569.75 & 113.95 \\
		  & 200 & 201.10 & 125.75 & 627.00 & 125.40 \\
		  & 220 & 212.20 & 135.25 & 673.75 & 134.75 \\
		  & 240 & 235.55 & 151.50 & 754.50 & 150.90 \\
		  & 260 & 252.45 & 160.30 & 798.00 & 159.60 \\
		  & 280 & 264.00 & 170.85 & 851.00 & 170.20 \\
		  & 300 & 311.40 & 194.60 & 970.50 & 194.10 \\
		\specialrule{.5pt}{0pt}{0pt}
	\end{tabular}
\end{table}
\begin{table}[bh!]
	\centering
	\caption{Results for average node degree $\deg = 4$, partition sizes $n \in \{3, 4, 5\}$ and node numbers $|V|=\{20,40,\ldots, 300\}$ for optimal and maximal $n$-soft domatic partitions. Showing the mean of the number of missing coverages $\overline{e_\text{miss\_cov}}$ and number of incompletely covered nodes $\overline{e_\text{inc\_nodes}}$.}
	\label{tbl:opt_max_4}
	\scriptsize
	\rowcolors{2}{gray!25}{white}
	\begin{tabular}{rr|rr|rr}
		\specialrule{.5pt}{0pt}{.5pt}
		\raggedleft $n$ &
		\raggedleft $|V|$ &
		\raggedleft $\overline{e_\text{miss\_cov}}$ &
		\raggedleft $\overline{e_\text{inc\_nodes}}$ &
		\raggedleft $\overline{e_\text{miss\_cov}}$ &
		\raggedleft $\overline{e_\text{inc\_nodes}}$ \tabularnewline
		\specialrule{.5pt}{.5pt}{.5pt}
		  &     & \multicolumn{2}{c|}{opt $n$-soft domatic partition} & \multicolumn{2}{c}{max $n$-soft domatic partition}\\
		3 &  20 &   0.75 &   0.75 &   2.25 &   0.75 \\
		  &  40 &   1.20 &   1.20 &   3.60 &   1.20 \\
		  &  60 &   1.50 &   1.50 &   4.50 &   1.50 \\
		  &  80 &   2.35 &   2.35 &   7.05 &   2.35 \\
		  & 100 &   2.25 &   2.25 &   6.75 &   2.25 \\
		  & 120 &   3.25 &   3.25 &   9.75 &   3.25 \\
		  & 140 &   3.20 &   3.20 &   9.60 &   3.20 \\
		  & 160 &   3.75 &   3.75 &  11.25 &   3.75 \\
		  & 180 &   3.70 &   3.70 &  11.10 &   3.70 \\
		  & 200 &   4.45 &   4.45 &  13.35 &   4.45 \\
		  & 220 &   3.80 &   3.80 &  11.40 &   3.80 \\
		  & 240 &   5.10 &   5.10 &  15.30 &   5.10 \\
		  & 260 &   4.70 &   4.70 &  14.10 &   4.70 \\
		  & 280 &   4.95 &   4.95 &  14.85 &   4.95 \\
		  & 300 &   6.50 &   6.50 &  19.50 &   6.50 \\
		4 &  20 &   4.00 &   3.25 &  13.00 &   3.25 \\
		  &  40 &   7.75 &   6.55 &  26.20 &   6.55 \\
		  &  60 &   9.05 &   7.55 &  30.20 &   7.55 \\
		  &  80 &  13.35 &  11.00 &  43.80 &  10.95 \\
		  & 100 &  15.25 &  13.00 &  51.80 &  12.95 \\
		  & 120 &  18.25 &  14.95 &  59.20 &  14.80 \\
		  & 140 &  19.50 &  16.30 &  65.20 &  16.30 \\
		  & 160 &  24.75 &  21.00 &  84.00 &  21.00 \\
		  & 180 &  23.40 &  19.70 &  78.60 &  19.65 \\
		  & 200 &  26.85 &  22.40 &  89.40 &  22.35 \\
		  & 220 &  26.85 &  23.00 &  92.00 &  23.00 \\
		  & 240 &  33.70 &  28.65 & 114.60 &  28.65 \\
		  & 260 &  33.10 &  28.40 & 113.60 &  28.40 \\
		  & 280 &  34.75 &  29.80 & 118.80 &  29.70 \\
		  & 300 &  41.90 &  35.40 & 141.40 &  35.35 \\
		5 &  20 &  11.70 &   7.65 &  38.25 &   7.65 \\
		  &  40 &  21.85 &  14.05 &  70.25 &  14.05 \\
		  &  60 &  30.25 &  21.15 & 105.25 &  21.05 \\
		  &  80 &  40.85 &  27.50 & 136.50 &  27.30 \\
		  & 100 &  47.75 &  32.50 & 162.25 &  32.45 \\
		  & 120 &  58.15 &  40.15 & 199.75 &  39.95 \\
		  & 140 &  63.45 &  44.15 & 220.25 &  44.05 \\
		  & 160 &  77.00 &  52.30 & 260.75 &  52.15 \\
		  & 180 &  80.15 &  56.95 & 283.00 &  56.60 \\
		  & 200 &  88.90 &  62.05 & 309.25 &  61.85 \\
		  & 220 &  93.15 &  66.50 & 331.25 &  66.25 \\
		  & 240 & 111.75 &  77.95 & 388.75 &  77.75 \\
		  & 260 & 114.65 &  81.65 & 406.25 &  81.25 \\
		  & 280 & 120.25 &  85.95 & 434.25 &  86.85 \\
		  & 300 & 141.90 & 100.35 & 499.50 &  99.90 \\
		\specialrule{.5pt}{0pt}{0pt}
	\end{tabular}
\end{table}
\begin{table}[bh!]
	\centering
	\caption{Results for average node degree $\deg = 5$, partition sizes $n \in \{3, 4, 5\}$ and node numbers $|V|=\{20,40,\ldots, 300\}$ for optimal and maximal $n$-soft domatic partitions. Showing the mean of the number of missing coverages $\overline{e_\text{miss\_cov}}$ and number of incompletely covered nodes $\overline{e_\text{inc\_nodes}}$.}
	\label{tbl:opt_max_5}
	\scriptsize
	\rowcolors{2}{gray!25}{white}
	\begin{tabular}{rr|rr|rr}
		\specialrule{.5pt}{0pt}{.5pt}
		\raggedleft $n$ &
		\raggedleft $|V|$ &
		\raggedleft $\overline{e_\text{miss\_cov}}$ &
		\raggedleft $\overline{e_\text{inc\_nodes}}$ &
		\raggedleft $\overline{e_\text{miss\_cov}}$ &
		\raggedleft $\overline{e_\text{inc\_nodes}}$ \tabularnewline
		\specialrule{.5pt}{.5pt}{.5pt}
		  &     & \multicolumn{2}{c|}{opt $n$-soft domatic partition} & \multicolumn{2}{c}{max $n$-soft domatic partition}\\
		3 &  20 &   0.20 &   0.20 &   0.60 &   0.20 \\
		  &  40 &   0.50 &   0.50 &   1.50 &   0.50 \\
		  &  60 &   0.60 &   0.60 &   1.80 &   0.60 \\
		  &  80 &   0.75 &   0.75 &   2.25 &   0.75 \\
		  & 100 &   1.05 &   1.05 &   3.15 &   1.05 \\
		  & 120 &   1.11 &   1.11 &   3.15 &   1.05 \\
		  & 140 &   1.35 &   1.35 &   4.05 &   1.35 \\
		  & 160 &   0.95 &   0.95 &   2.85 &   0.95 \\
		  & 180 &   1.40 &   1.40 &   4.20 &   1.40 \\
		  & 200 &   1.25 &   1.25 &   3.75 &   1.25 \\
		  & 220 &   1.40 &   1.40 &   4.20 &   1.40 \\
		  & 240 &   1.60 &   1.60 &   4.80 &   1.60 \\
		  & 260 &   1.75 &   1.75 &   5.25 &   1.75 \\
		  & 280 &   1.75 &   1.75 &   5.25 &   1.75 \\
		  & 300 &   2.00 &   2.00 &   6.00 &   2.00 \\
		4 &  20 &   1.75 &   1.55 &   6.20 &   1.55 \\
		  &  40 &   3.35 &   2.85 &  11.40 &   2.85 \\
		  &  60 &   4.25 &   3.65 &  14.60 &   3.65 \\
		  &  80 &   6.15 &   5.40 &  21.60 &   5.40 \\
		  & 100 &   6.80 &   5.75 &  23.00 &   5.75 \\
		  & 120 &   6.75 &   5.70 &  22.80 &   5.70 \\
		  & 140 &   8.15 &   6.80 &  26.31 &   6.58 \\
		  & 160 &   8.55 &   7.60 &  30.40 &   7.60 \\
		  & 180 &  10.30 &   8.90 &  35.60 &   8.90 \\
		  & 200 &  10.25 &   9.00 &  36.00 &   9.00 \\
		  & 220 &  10.65 &   9.25 &  37.00 &   9.25 \\
		  & 240 &  12.15 &  10.55 &  42.20 &  10.55 \\
		  & 260 &  13.40 &  11.65 &  46.60 &  11.65 \\
		  & 280 &  14.85 &  13.10 &  52.40 &  13.10 \\
		  & 300 &  15.35 &  13.35 &  53.40 &  13.35 \\
		5 &  20 &   5.80 &   4.05 &  20.25 &   4.05 \\
		  &  40 &  11.15 &   7.80 &  39.00 &   7.80 \\
		  &  60 &  15.40 &  11.15 &  55.75 &  11.15 \\
		  &  80 &  19.80 &  13.65 &  68.25 &  13.65 \\
		  & 100 &  23.60 &  16.85 &  84.25 &  16.85 \\
		  & 120 &  26.20 &  19.45 &  97.25 &  19.45 \\
		  & 140 &  29.10 &  20.90 & 104.25 &  20.85 \\
		  & 160 &  31.40 &  22.75 & 113.50 &  22.70 \\
		  & 180 &  38.90 &  28.70 & 143.50 &  28.70 \\
		  & 200 &  40.45 &  30.20 & 151.00 &  30.20 \\
		  & 220 &  43.05 &  32.40 & 161.50 &  32.30 \\
		  & 240 &  47.80 &  35.65 & 178.00 &  35.60 \\
		  & 260 &  52.10 &  38.60 & 193.00 &  38.60 \\
		  & 280 &  54.55 &  39.60 & 198.00 &  39.60 \\
		  & 300 &  59.25 &  43.80 & 219.00 &  43.80 \\
		\specialrule{.5pt}{0pt}{0pt}
	\end{tabular}
\end{table}
\begin{table}[bh!]
	\centering
	\caption{Results for average node degree $\deg = 6$, partition sizes $n \in \{3, 4, 5\}$ and node numbers $|V|=\{20,40,\ldots, 300\}$ for optimal and maximal $n$-soft domatic partitions. Showing the mean of the number of missing coverages $\overline{e_\text{miss\_cov}}$ and number of incompletely covered nodes $\overline{e_\text{inc\_nodes}}$.}
	\label{tbl:opt_max_6}
	\scriptsize
	\rowcolors{2}{gray!25}{white}
	\begin{tabular}{rr|rr|rr}
		\specialrule{.5pt}{0pt}{.5pt}
		\raggedleft $n$ &
		\raggedleft $|V|$ &
		\raggedleft $\overline{e_\text{miss\_cov}}$ &
		\raggedleft $\overline{e_\text{inc\_nodes}}$ &
		\raggedleft $\overline{e_\text{miss\_cov}}$ &
		\raggedleft $\overline{e_\text{inc\_nodes}}$ \tabularnewline
		\specialrule{.5pt}{.5pt}{.5pt}
		  &     & \multicolumn{2}{c|}{opt $n$-soft domatic partition} & \multicolumn{2}{c}{max $n$-soft domatic partition}\\
		3 &  20 &   0.10 &   0.10 &   0.31 &   0.11 \\
		  &  40 &   0.15 &   0.15 &   0.45 &   0.15 \\
		  &  60 &   0.30 &   0.30 &   0.90 &   0.30 \\
		  &  80 &   0.30 &   0.30 &   0.90 &   0.30 \\
		  & 100 &   0.20 &   0.20 &   0.60 &   0.20 \\
		  & 120 &   0.35 &   0.35 &   1.05 &   0.35 \\
		  & 140 &   0.50 &   0.50 &   1.50 &   0.50 \\
		  & 160 &   0.45 &   0.45 &   1.35 &   0.45 \\
		  & 180 &   0.40 &   0.40 &   1.20 &   0.40 \\
		  & 200 &   0.80 &   0.80 &   2.40 &   0.80 \\
		  & 220 &   0.85 &   0.85 &   2.55 &   0.85 \\
		  & 240 &   0.80 &   0.80 &   2.40 &   0.80 \\
		  & 260 &   0.90 &   0.90 &   2.70 &   0.90 \\
		  & 280 &   0.90 &   0.90 &   2.70 &   0.90 \\
		  & 300 &   0.80 &   0.80 &   2.40 &   0.80 \\
		4 &  20 &   0.75 &   0.65 &   2.60 &   0.65 \\
		  &  40 &   1.55 &   1.40 &   5.60 &   1.40 \\
		  &  60 &   2.15 &   1.85 &   7.40 &   1.85 \\
		  &  80 &   2.35 &   2.05 &   8.20 &   2.05 \\
		  & 100 &   2.70 &   2.50 &  10.00 &   2.50 \\
		  & 120 &   3.50 &   3.15 &  12.60 &   3.15 \\
		  & 140 &   3.65 &   3.15 &  12.60 &   3.15 \\
		  & 160 &   3.70 &   3.25 &  13.00 &   3.25 \\
		  & 180 &   4.10 &   3.70 &  14.80 &   3.70 \\
		  & 200 &   4.55 &   3.75 &  15.00 &   3.75 \\
		  & 220 &   5.45 &   4.60 &  18.40 &   4.60 \\
		  & 240 &   5.45 &   4.65 &  18.60 &   4.65 \\
		  & 260 &   6.15 &   5.25 &  21.00 &   5.25 \\
		  & 280 &   6.20 &   5.30 &  21.20 &   5.30 \\
		  & 300 &   6.85 &   6.05 &  24.20 &   6.05 \\
		5 &  20 &   2.75 &   2.00 &  10.00 &   2.00 \\
		  &  40 &   6.45 &   4.90 &  24.50 &   4.90 \\
		  &  60 &   8.65 &   6.50 &  32.50 &   6.50 \\
		  &  80 &   9.80 &   7.50 &  37.50 &   7.50 \\
		  & 100 &  11.50 &   8.80 &  44.00 &   8.80 \\
		  & 120 &  14.30 &  10.80 &  54.00 &  10.80 \\
		  & 140 &  15.20 &  11.55 &  57.75 &  11.55 \\
		  & 160 &  14.90 &  11.20 &  56.00 &  11.20 \\
		  & 180 &  18.25 &  14.15 &  70.75 &  14.15 \\
		  & 200 &  18.70 &  14.20 &  71.00 &  14.20 \\
		  & 220 &  21.45 &  16.00 &  80.00 &  16.00 \\
		  & 240 &  23.90 &  18.45 &  92.25 &  18.45 \\
		  & 260 &  23.25 &  17.10 &  85.50 &  17.10 \\
		  & 280 &  26.10 &  19.95 &  99.75 &  19.95 \\
		  & 300 &  28.40 &  21.55 & 107.75 &  21.55 \\
		\specialrule{.5pt}{0pt}{0pt}
	\end{tabular}
\end{table}
Even so, we compare results which have been computed for different graphs and for each parameter combination, we set up only a set of $20$ graphs.
For each row in the Tables \ref{tbl:opt_max_3}, \ref{tbl:opt_max_4}, \ref{tbl:opt_max_5} and \ref{tbl:opt_max_6} and its respective parameter combinations, we generated $20$ graphs and determined the optimal and maximal $3$, $4$ and $5$-soft domatic partitions. 
As the data used in the plots mirror the information showcased in the tables, all presented data signify optimal solutions identified through Gurobi within the time limit of $1200$ seconds.
Small fluctuations in the results can be caused by Gurobi even without a given MIPGap. 
By default, Gurobi aims to prove optimality within certain numerical tolerances without the user explicitly setting the MIPGap parameter. 
Gurobi's MIPGap represents the allowable gap between the best-known solution and the proven optimal solution.
Therefore, in some cases, we can observe those numerical tolerances in the table.
To compare the results of the maximal and optimal $n$-soft domatic partitions, we evaluate Fig. \ref{fig:inc_cov_max_opt} in which we reflect the number of incompletely covered nodes as result of the computation of the maximal and optimal $n$-soft domatic partition.
The dashed and dotted lines in the diagrams serve solely to enhance readability and do not correspond to computed data or specific interpretations.

For a final comparison of the performance of the solutions for maximal and optimal $n$-soft domatic partitions, we determine the relative mean of the results for the graphs in $SG_1$.
On average, the number of missing coverages $e_\text{miss\_cov}$ of the optimal $n$-soft domatic partition for our test setup and for the set of graphs $SG_1$ is $10.52\%$ lower compared to the maximal $n$-soft domatic partition.  
In contrast, the same comparison for the number of incompletely covered nodes yields only an improvement of $\overline{P_\text{inc\_nodes}}=0.04\%$ for the maximal compared to the optimal $n$-soft domatic partition. 
Meaning that on average the number of incompletely covered nodes for the maximal $n$-soft domatic partition is $0.04\%$ lower compared to the optimal $n$-soft domatic partition. 
In Tables \ref{tbl:opt_max_3}, \ref{tbl:opt_max_4}, \ref{tbl:opt_max_5} and \ref{tbl:opt_max_6} absolute values are shown that give an impression on the behaviour of the number of incompletely covered nodes and the number of missing coverages resulting from the maximal and optimal $n$-soft domatic partitions.

For graphs in $SG_2$, we adapted the graphs from $SG_1$ to be bridge-free. 
Our expectation was that this property has a significant impact on the computation time and on the quality of results regarding the number of missing coverages and incompletely covered nodes. 
Despite our expectations, the results yield that there exists no notable difference between the quality of results and the computation time. 
Our simulation case studies demonstrate that the elimination of bridges does not imply a measurable effect on the computation time necessary to obtain optimal and maximal $n$-soft domatic partitions. 

For all graphs, we have been able to compute the optimal and maximal $n$-soft domatic partitions to optimality. 
Moreover, our results illustrated that the optimal $n$-soft domatic partitions exhibit an almost similar number of incompletely covered nodes as the maximal $n$-soft domatic partition. 
The number of missing coverages on the other hand increases significantly for the maximal $n$-soft domatic partitions in contrast to the optimal $n$-soft domatic partitions.
The computation time seems to increase significantly with an rising average node degree.
All together, the tests revealed that for all considered large-scale static homogeneous WSNs, the computation of maximal and optimal $n$-soft domatic partitions is possible and yields an optimal solution.

\section{Application}\label{sec:application}
There are multiple applications for our partitioning scheme to contribute to the security of large-scale static homogeneous WSNs.
Our partitioning schemes facilitate an equitable distribution of security means, aiming to ensure the availability of $n$ distinct security means in close proximity to each node.
The selection of a security configuration is highly contingent upon the application area (topology, environmental conditions, accessibility for potential attackers), the specific security requirements, and other factors such as sensor node hardware and network lifespan.
When designing and practically implementing a security framework which combines an ensemble of security features it is essential to assess possible emerging vulnerabilities.
Here, we are drawing a rough picture on how to utilise our proposed partitioning schemes for a straightforward ensemble security framework.
Hence, we will not examine all the implementation details and skip a comprehensive security analysis to determine arising vulnerabilities.
We contemplate a large-scale static WSN deployed in a forest for the purpose of environmental monitoring, comprising several hundred nodes.
The data's confidentiality within this network is not paramount due to the low sensitivity of individual measurement data.
However, the integrity and authenticity of the data is crucial for detecting dangers to the ecosystem and potentially extreme events (e.g., wildfires).
Therefore, we create an area of application and elucidate the choice of the two security means to be distributed in the WSN contributing to those requirements.
We outline a combination of intrusion detection with an agent-based rerouting concept \cite{tomic2018antilizer} and an information hiding scheme (invisible watermarking/steganography) \cite{evsutin2022overview,wang2016information,de2014applying,sultana2012secure,zander2007survey}. 
The \textit{Antilizer}, as proposed in \cite{tomic2018antilizer} is a network-level IDS and automated trust-based response system.
It uses an agent-based notification (ANT) scheme to detect malicious behaviour.
Therefore, each node builds a trust-model of its neighbours.
The trust-model is used to make routing decisions at each node.
The ANTs are sent to the BS and notify intermediate nodes along the way that routing changes are the result of malicious behaviour.
To detect those changes in the first place, nodes overhear their one-hop neighbours.
Each node self-collects information about transmissions, receptions and further communication events ignoring the untrusted message content.
Additionally, the trust-model can be adjusted by responses of the BS.
The corresponding routing decisions are then made using the trust model.
The \textit{Antilizer} contributes to the node integrity and authenticity by detecting malicious behaviour and notifying the BS.
As reaction, the BS utilises a filtering mechanism to determine the validity of notifications. 
We combine the \textit{Antilizer} scheme with an invisible watermarking scheme.
The combination implicates that not all nodes maintain their own trust-model and therefore, are required to rely on trust-models of neighbouring nodes.

To justify our choice of an information hiding scheme, we need to delve into the necessary background and provide some context.
We consider a lightweight information hiding scheme with focus on data integrity and the protection of the source of origin of transmitted data (authenticity).
Suitable concepts are fragile invisible watermarking or steganographic information hiding schemes \cite{evsutin2022overview}.
We deem a fragile scheme to be satisfactory, assuming that the absence of a watermark serves as a sufficient indication of tampered data.

We propose to utilise pseudo-image watermarking to allow aggregation of data \cite{xiao2019digital,evsutin2022overview}.
Already watermarked data, presented as pseudo-images, that traverse through an additional watermarking/aggregation node are exclusively forwarded to safeguard the fragile watermark.
Alternatively, an aggregation tolerant watermark can be considered \cite{panah2015shadows}.
To reach the critical amount of data necessary to create a pseudo-image, either forwarded data of other nodes or, in the case of insufficient forwarded data, the node's own data are collected and aggregated.
The temporal accumulation of own data potentially necessitates the usage of time codes or an order as meta information for the evaluation at the BS.
The watermarking scheme we propose does not yet exist in the literature, but its components and their integration to some extent do \cite{xiao2019digital}.
Modifications to certain levels are imperative to ensure their applicability in WSNs, where only a subset of nodes employ them. 

Using our partitioning scheme, we distribute the proposed security means in a WSN.
We compute the $n$-soft domatic partition of nodes for partition sizes $n$ of $2$ and $3$.
The sets in a partition can differ in size affecting the ratio of different security means associated with them.
Further, the union of dominating sets of a partition results in a dominating set.
We can use this property to control the ratio of nodes implementing selected security means.
Even so, we intend to distribute two types of security means, if a partition size of $3$ is achievable with a small number of missing coverages, it potentially provides a certain amount of control over the ratio between watermarking/aggregation nodes and \textit{Antilizer} nodes.
Further, we provided auxiliary tools in the preceding sections to adjust the set sizes in $n$-soft domatic partitions.
Under certain assumptions, it is helpful to set selected nodes, e.g. nodes on the outer rim of a WSN, to a specific security mean.
Such adjustments can be considered in the partitioning by constraining the corresponding variables to a specific set of a partition and computing the remaining variables accordingly.
In our case, the nodes on the outer rim can potentially be bound to the watermarking/aggregation scheme.
We assume that outer nodes are more prone to attacks in the considered scenario.
Another positive effect is the limited frequency of data transmissions, since those nodes are assumed to aggregate mostly own data.
Further, the watermarking allows an early detection of attacks on the data integrity.
Subsequently, the nodes utilising \textit{Antilizer} would be placed on inner nodes of the network, enabling them to potentially overhear an increased number of nodes.

In general, nodes are gathering data and forwarding them hop-by-hop to a BS.
The BS validates the data by comparing measurements with past measurement data as well as the variance of data in proximity.
To have reliable data in the proximity of nodes, we watermark data at certain nodes and mark the data collected by those nodes specifically.
The presence/absence of a watermark is validated by the BS.
Since, we expect measurement data to be similar in local proximity (or their respective gradients are smooth to some degree), the BS validates the data by comparing them to watermarked measurements.
While missing watermarks lead to a decrease in reputation of nodes, the presence can restore or maintain a level of reputation.
Parallel to the watermarking scheme, the \textit{Antilizer} nodes provide a timely threat reaction by monitoring network-level behaviour, adjusting the reputation of neighbouring nodes and influence routing decisions based on those.
Those information is passed by agents (ANTs) to the BS for further evaluation.
The BS can then create an overarching picture of network metric changes and measurement deviations to determine suitable adjustments.
The evaluation result of the BS can be used to adjust trust-models of nodes.
Routing decisions of nodes not implementing \textit{Antilizer} rely on BS responses in combination with trust-models determined by neighbouring nodes. 

With limited resources, it is impossible to fend off all possible attacks and to achieve perfect security.
However, the combination of security means can strengthen various security features and provide resilience against a certain type of attackers.
In general, we know from nature that versatility is a key to a strong and efficient, but imperfect security \cite{forster2023novel}.

\section{Conclusion and Future Prospects}\label{sec:conclusion}
In this paper, we determined a distribution of security means based on the concept of a neighbourhood watch introduced by Langendörfer \cite{langend2019security}. 
The concept aims to maximise the spectrum of security threats a large-scale static homogeneous WSN can detect or avert while minimising the load that will be put on individual nodes. 
To develop a complex security framework of this kind, there are several steps that have to be taken. 
Here, we introduced a graph partitioning scheme for the node distribution. 
While sleep scheduling themes allow partitioning schemes that determine non-disjoint minimal dominating sets, we were looking for a partition that creates disjoint partitions that approximate the definition of dominating sets. 
Therefore, we defined two terms, the number of missing coverages and the number of incompletely covered nodes. 
To determine the partitions based on those terms, we introduced two $0-1$ LPs for the maximal $n$-soft domatic partition and for the optimal $n$-soft domatic partition. 
Furthermore, we proposed several variations of those LPs allowing advanced distributions of security means that fit to the needs of differently equipped WSNs and to different levels of security threats.
To validate the computability of the proposed NP hard $0-1$ LPs, a test setup has been designed.
On its basis, we have verified the computability on graphs as representations of large-scale static homogeneous WSNs.
This also implied the need for a suitable graph generator that enables to create realistic WSN models. 
The introduced graph generator allows to control the properties of resulting graphs via its input parameters, allowing an improved comparability of test results. 
Further, our graph generator aims at the creation of connected graphs as far as possible by purposive construction from the beginning. 
This feature avoids expensive trial-and-error strategies by iterating over a large number of insufficient graphs. 
Along with algorithmic design, we had to cope with the requirement that the constructive generation of connected graphs does not interfere with the desired uniform node distribution.
As a result, we developed a new graph generator for $\lambda$-precision UDGs introduced in this publication.
Its Python source code is available from the first author upon request.
Additionally, further major properties we are able to control to some extent are the average node degree, the local clustering coefficient and the general coverage of the generation plane.
Beyond, we provide several methods to further adapt the resulting graphs while maintaining their characteristics as representations of WSNs.

To evaluate the introduced LPs, we introduced a generator for $\lambda$-precision UDGs. 
The generator enabled us to evaluate which parameters affect the computation time at most by providing appropriate graphs.
Our results have shown that the computation time is affected the most by the average node degree and the desired partition size. 
In the range of $20$ to $300$ nodes within a graph, the node degree has almost only a linear effect on the computation time.
The application in the last section intends to visualise the generic applicability of our partitioning scheme for the timely efficient design of complex and cooperative security configurations.

We have presented a number of variations towards the $0-1$ LPs allowing distributions of a fixed number of security means per node and even distributions based on the performance cost of each security mean. 
The latter one allows to distribute varying numbers of security means per node based on their individual resource requirements. 
Our future work will address a number of applications with the goal to design a DSE framework for requirement-based cooperative/collaborative security configurations for WSNs.
The process can conclude by tailoring the chosen security means to align with the requirements of a specific WSN, limiting the size of the design space as outlined in \cite{langend2019security}, utilising the proposed CADRT.
Our graph generator is suited for adaptation to determination of automatic node distributions for given topologies.
Those adaptations include the consideration of obstacles and elevation profiles as well as node capabilities.
The current version of the $\lambda$-precision UDG generator is effortless adjustable to arbitrary formed areas, automatically identifying appropriate uniform node distributions to achieve desired coverages.
Only if desired coverages for a set of given sensor nodes and its capabilities are technically not achievable, manual intervention is necessary, e.g. adjusting the size of the given distribution area.

\bibliographystyle{unsrt}
\bibliography{bib}

\begin{IEEEbiography}[{\includegraphics[width=1in,height=1.25in,clip,keepaspectratio]{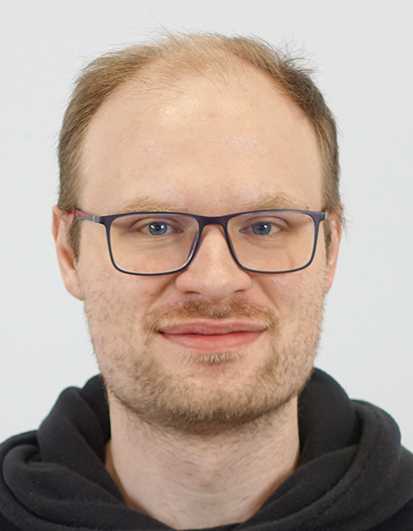}}]{Benjamin Förster} was born in Frankfurt (Oder), Germany in 1991 and received his B.Sc. and M.Sc. in Computer Science at Brandenburg University of Technology Cottbus-Senftenberg (B-TU), in Cottbus, Brandenburg, Germany in 2017 and 2020 respectively.

	He is currently a Ph.D. Student at IHP in Frankfurt (Oder), Germany. His major field of study is security in wireless embedded resource constraint systems. Prior he worked as research assistant at IHP and B-TU and software developer at Astronergy Solarmodule GmbH.
	His research interests encompass a range of topics, including the security of wireless resource-constrained embedded systems and the application of biologically inspired computing principles.
\end{IEEEbiography}
\begin{IEEEbiography}[{\includegraphics[width=1in,height=1.25in,clip,keepaspectratio]{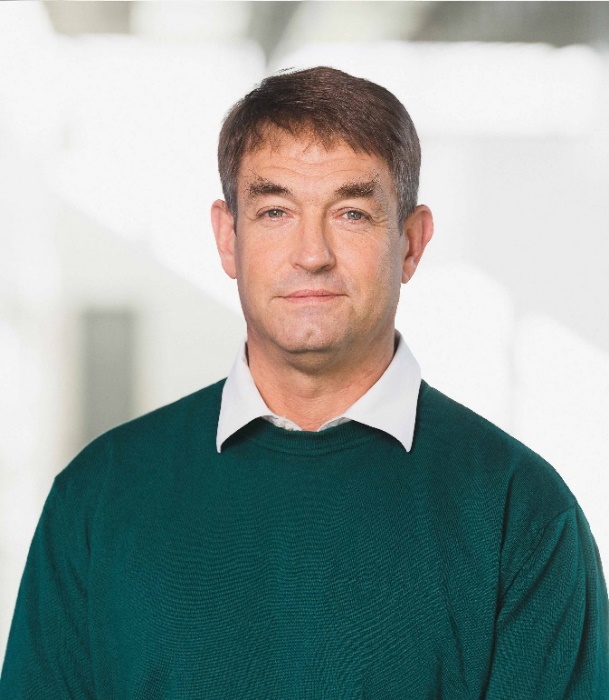}}]{Peter Langendörfer} holds a diploma and a doctorate degree in computer science from B-TU in Cottbus, Brandenburg, Germany received in 1995 and 2001 respectively.

	He is with the IHP in Frankfurt (Oder) since 2000. There, he is leading the wireless systems department.
	From 2012 till 2020 he was leading the chair for security in pervasive systems at the Technical University of Cottbus-Senftenberg. 
	Since 2020 he owns the chair wireless systems at the Technical University of Cottbus-Senftenberg.
	His portfolio includes over 150 peer-reviewed technical articles that he has published, as well as 17 filed patents, with 11 already granted.
	He is associate editor of IEEE Access, Peer-to-Peer Networking and worked as guest editor for many renowned journals e.g. Wireless Communications and Mobile Computing (Wiley) and ACM Transactions on Internet Technology.
	His areas of high interest include security for resource constraint devices, low-power protocols, efficient implementations of AI, and resilience.

	Prof. Dr. Langendörfer is member of the ``Gesellschaft für Informatik''.
\end{IEEEbiography}
\begin{IEEEbiography}[{\includegraphics[width=1in,height=1.25in,clip,keepaspectratio]{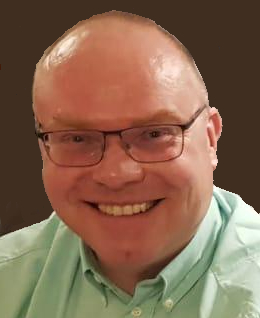}}]{Thomas Hinze} was born in Halle (Saale), Germany in 1971. He received the Diploma degree in computer science from the Dresden University of Technology, Germany in 1997 and the Ph.D. degree in engineering sciences from the same institution in 2002. In 2012, he became a senior university lecturer at the Friedrich Schiller University Jena, Germany, along with his professorial dissertation.

	He is author of three text books, more than 90 publications and holds two patents. His research interests include principles of biological and biologically inspired information processing like molecular computing, membrane computing, distributed computing, evolutionary computing, and systems biology. He is an editorial board member of the Springer Journal of Membrane Computing.

	PD Dr. Hinze joined the Association for Computing Machinery (ACM) and engages in the International Membrane Computing Society (IMCS) as well as in the Institute for Systems and Technologies of Information, Control and Communication (INSTICC). He was a recipient of five Best Paper Awards.
\end{IEEEbiography}

\EOD

\end{document}